\documentclass[reprint,eqsecnum,floats,aps,amsmath,amssymb,nofootinbib,prd,onecolumn, showpacs]{revtex4-1}

\usepackage{graphicx}
\usepackage{amsmath,amssymb}
\usepackage{hyperref}
\usepackage{graphicx}
\usepackage{subfigure}
\usepackage{arydshln}

\begin{document}

\title{Fermions in Hybrid Loop Quantum Cosmology}

\author{Beatriz Elizaga Navascu\'es}
\email{beatriz.elizaga@iem.cfmac.csic.es}
\affiliation{Instituto de Estructura de la Materia, IEM-CSIC, Serrano 121, 28006 Madrid, Spain}
\author{Mercedes Mart\'{\i}n-Benito}
\email{mmartinb@fc.ul.pt}
\affiliation{Instituto de Astrof\'isica e Ci\^encias do Espa\c co, Universidade de Lisboa, Faculdade de Ci\^encias, Ed. C8, Campo Grande, PT1749-016 Lisboa, Portugal}
\author{Guillermo  A. Mena Marug\'an}
\email{mena@iem.cfmac.csic.es}
\affiliation{Instituto de Estructura de la Materia, IEM-CSIC, Serrano 121, 28006 Madrid, Spain}

\begin{abstract}
This work pioneers the quantization of primordial fermion perturbations in hybrid Loop Quantum Cosmology (LQC). We consider a Dirac field coupled to a spatially flat, homogeneous, and isotropic cosmology, sourced by a scalar inflaton, and treat the Dirac field as a perturbation. We describe the inhomogeneities of this field in terms of creation and annihilation variables, chosen to admit a unitary evolution if the Dirac fermion were treated as a test field. Considering instead the full system, we truncate its action at quadratic perturbative order and construct a canonical formulation. In particular this implies that, in the global Hamiltonian constraint of the model, the contribution of the homogeneous sector is corrected with a quadratic perturbative term. We then adopt the hybrid LQC approach to quantize the full model, combining the loop representation of the homogeneous geometry with the Fock quantization of the inhomogeneities. We assume a Born-Oppenheimer ansatz for physical states and show how to obtain a Schr\"odinger  equation for the quantum evolution of the perturbations, where the role of time is played by the homogeneous inflaton. We prove that the resulting quantum evolution of the Dirac field is indeed unitary, despite the fact that the underlying homogeneous geometry has been quantized as well. Remarkably, in such evolution, the fermion field couples to an infinite sequence of quantum moments of the homogeneous geometry. Moreover, the evolved Fock vacuum of our fermion perturbations is shown to be an exact solution of the Schr\"odinger equation. Finally, we discuss in detail the quantum backreaction that the fermion field introduces in the global Hamiltonian constraint. For completeness, our quantum study includes since the beginning (gauge-invariant) scalar and tensor perturbations, that were studied in previous works.
\end{abstract}



\maketitle

\section{Introduction}

Observational cosmology has gone through impressive developments in the last decades, with improvements in the resolution that, in particular, have allowed us to determine several cosmological parameters with an error of a few percent \cite{planck,planck-inf}. In this new era of precision cosmology, the Cosmic Microwave Background (CMB) is an important source of information about the physics of the Early Universe and the primordial seeds for the later formation of structures \cite{liddle}. The observation of the CMB has given a solid support to the idea that those seeds originated from quantum fluctuations of the perturbations of a flat, homogeneous, and isotropic state of the Universe that underwent a period of inflation \cite{mukh,langlois}. The most conventional approach to describe the intervening inflationary mechanism is to introduce a scalar field (the inflaton) subject to a potential, the contribution of which drives the expansion. The consideration of primordial perturbations of scalar and tensor nature within the inflationary paradigm leads to great agreement between theoretical and observational predictions.
Nonetheless, most of the fundamental theories for matter interactions involve nonbosonic degrees of freedom, and in particular spin-$1/2$ ones. Thus, it is interesting to incorporate as well this type of fields in order to discuss the physics of the Early Universe in a more realistic manner. This discussion would elucidate whether the presence and evolution of fermion fields during the first cosmological stages may have left any trace in the scalar and tensor primordial perturbations, and investigate quantitatively the extent to which they did or did not not affect them.

Physical effects of free fermionic fields viewed as perturbations propagating in a flat cosmology, both during inflation and in other epochs of the Universe, have been discussed in several works, treating them as test fields in the framework of quantum field theory \cite{cosmof1,cosmof2,cosmof3,cosmof4,cosmof5,cosmof6,cosmof7,cosmof8}. The first analyses that studied fermions in quantum cosmology, that is, assuming that the cosmological homogeneous background is as well a quantum entity, can be traced back to the 70's \cite{isham,christo}. Interestingly, these works did not regard the fermionic matter as a test field but rather as the source of geometry. However, they restricted the fermionic field to be homogeneous as well, therefore reducing its degrees of freedom to a finite number. 

In the context of quantum geometrodynamics \cite{halliwell}, the inclusion of fermions was made by D'Eath and Halliwell in a study particularized to Dirac spinors \cite{DEH}. They extended to these fields the treatment of cosmological perturbations in a spatially closed universe put forward by Halliwell and Hawking a few years before \cite{HH}. In their paper, D'Eath and Halliwell adopted a holomorphic representation for the fermionic field, and introduced creation and annihilation variables that lead to an instantaneous diagonalization of the Hamiltonian. They argued that the production of the corresponding particles is finite, and discussed issues related to the backreaction of the fermions onto the quantum background and perturbations of the geometry and  the inflaton.
 
The main purpose of the present work is to generalize this treatment of fermionic fields in quantum cosmology in order to allow for other approaches to the quantization of the background instead of quantum geometrodynamics, with our emphasis placed on the loop quantization program \cite{lqc1}. Besides, we will also use recent results on the Fock representation of fermions \cite{uf1,uf2,uf-flat} with the aim at improving some properties of the quantization of the Dirac field related with the unitarity of the evolution and the backreaction effects. Owing to its physical relevance, we will adopt a flat topology for the spatial sections of our model.

Loop Quantum Cosmology (LQC) \cite{lqc1,lqc2,lqc3,lqc4} is a quantization approach for cosmological spacetimes that is based on the methods of Loop Quantum Gravity (LQG) \cite{lqg}. This latter formalism is one of the most solid candidates to construct a nonperturbative quantum theory of gravity. LQG is a canonical program for the quantization of general relativity that is independent of background structures. The application of LQC to homogeneous and isotropic spacetimes has led to remarkable results, among which the most celebrated one is the resolution of the cosmological singularity, that is replaced by a quantum bounce \cite{APS1,APS2,MMO,singh}. 
The framework of LQC was enlarged to allow for the consideration of inhomogeneous spacetimes with the introduction of hybrid quantization methods \cite{hybr-rev}, based on combining the loop quantization of certain global modes of the geometry with other, more conventional quantization techniques for the additional inhomogeneities. This procedure, known with the name of hybrid LQC, was first applied to linearly polarized Gowdy cosmologies \cite{hybr-gow1,hybr-gow2,hybr-gow3,hybr-gow4}. It was soon extended to realistic cosmological scenarios corresponding to perturbed homogeneous and isotropic universes, both with scalar \cite{hybr-inf1,hybr-inf2,hybr-inf3,hybr-ref,hybr-inf5} and tensor perturbations \cite{hybr-ten}. Other methods have also been developed for the quantization of cosmological perturbations in LQC. This is the case, e.g., of the dressed metric approach \cite{dressed1,dressed2,dressed3,dressed4}, which employs as well hybrid quantization techniques, but renouncing to the description of the whole cosmological system as a symplectic constrained system. The consequences of the loop quantization of the geometry on the CMB have also been studied using effective equations derived from arguments related with the closure of the quantum constraint algebra and certain additional assumptions about the form of the quantum corrections \cite{effective1,effective2,effective3,effective4,effective5}. For a recent review of these different lines of attack, see Ref. \cite{Edward}. On the other hand, the effect of hybrid LQC on scalar and tensor perturbations, its reflection in the correlation functions of the CMB, and the comparison with observations, have been investigated in Refs. \cite{hybr-pred,hybr-lon}. The predictions extracted so far seem in good agreement with the observations. 

The consideration of fermions in LQC has been limited to the study of a homogeneous and anisotropic model coupled to a homogeneous fermion field \cite{boj-fermions}, where the focus was placed in the role played by parity. In this work we aim to analyzing a fermion field with local degrees of freedom in the context of hybrid LQC. The interest of considering fermions in hybrid LQC exceeds the inclusion of realistic fundamental matter fields in our quantum description of the Early Universe. On the one hand, the introduction of fermions puts to the test the very own consistency of the hybrid approach, requiring a quantization where the loop quantum geometry is coupled to fermionic fields. Thus, one faces the challenge of combining in a consistent way, in a constrained infinite-dimensional system, the polymeric representation of the geometry with a more conventional representation (e.g. a Fock representation) of appropriate canonical anticommutation relations for the fermions, following the rules of quantum field theory in curved spacetimes \cite{wald}. On the other hand, LQC provides a complete and fully controlled quantization of homogeneous and isotropic universes. Therefore, one expects that hybrid LQC will allow us to treat fermions in cosmology in a genuine quantum way, without recurring at any moment to semiclassical approximations, as it was the case in Ref. \cite{DEH}. This quantum treatment is important if one wants to investigate thoroughly physical processes that really belong to the quantum realm. Moreover, the fact that the polymeric quantization is inequivalent to other traditional quantizations (like geometrodynamics) casts doubts on whether one can ignore the effects on the cosmological evolution of the quantum fermionic fields. In particular, the existence of a bounce eliminates the singularity of the geometry, and may change drastically the behavior of the fermions, even when described in the context of quantum field theory in a nonstationary background. Partly related with this issue is the correct definition of a vacuum state. The corrections caused by the loop quantization alter the geometry, changing its dynamics and the corresponding symmetries of the spacetime in effective descriptions. It is then natural to expect that these changes modify as well the vacuum state, at least if one understands it as a state that is optimally adapted to the background dynamics. Another important issue is the backreaction that the quantum fermions produce onto the geometry. There is an increasing interest on this problem in LQC, and although some very preliminary discussions have been carried out in different contexts \cite{gow-back,hybr-num}, the development of fully self-consistent formalisms to study this backreaction is a necessary step before dealing successfully with it. These are the questions and problems that we want to discuss in this work.

The rest of the article is organized as follow. In Sec. II we will describe our classical system, that represents a flat, homogeneous, and isotropic Universe, with a homogeneous massive scalar field, that contains scalar and tensor perturbations, as well as a Dirac field that is also treated as a perturbation. In Sec. III we will discuss the choice of creation and annihilation variables for the Dirac field. We will summarize previous results on the uniqueness of this choice if one demands a unitary dynamics for the selected variables, together with some symmetry requirements. This criterion picks out a family of Fock representations that are all unitarily equivalent. We will complete the change that leads to such creation and annihilation variables, complementing it with a change of homogeneous variables. The total transformation performed in the system renders it canonical at the considered perturbative level. Although we will try to maintain our discussion as general as possible, independently of a specific choice in the privileged family of representations that is selected by our criteria, in Sec. IV we will focus our attention on the representation associated with the same creation and annihilation variables that were introduced by D'Eath and Halliwell, given the interest in comparing our results with those of Ref. \cite{DEH}. We will then proceed to the hybrid quantization of the resulting system in Sec. V. In Sec. VI we will adopt a Born-Oppenheimer ansatz that separates the dependence of the quantum states on the homogeneous geometry, the scalar and tensor perturbations, and the fermionic degrees of freedom. We will discuss conditions to arrive in this way to a Schr\"odinger equation for the fermionic part. In Sec. VII, we will study the quantum dynamics of our creation and annihilation operators for the nonzero-modes of the Dirac field. We will implement the dynamics by means of an evolution operator in Sec. VIII, where we will also construct solutions to the associated Schr\"odinger equation and provide a fermionic Hamiltonian. In Sec. IX we will prove the unitarity of the quantum dynamics and discuss the production of particles and the fermionic backreaction contribution. Finally, in Sec. X we conclude discussing our results. We use units such that the Newton constant, the speed of light, and the reduced Planck constant are equal to one.

\section{The model}

\subsection{The unperturbed homogeneous model}

We start by considering a Friedmann-Lema\^{\i}tre-Robertson-Walker (FLRW) spacetime in which the homogeneous and isotropic spatial sections have flat compact topology, so that they are isomorphic to a three-torus $T^3$. This assumption of compactness should be irrelevant for cosmological purposes if the compactification radius is sufficiently large (typically much larger than the associated Hubble radius), but will simplify considerably our mathematical treatment. The corresponding metric can be written in the form 
\begin{equation}
\label{homometric}
ds^2=\sigma^2 \big(-N_0^2(t) dt^2+e^{2\alpha(t)} \,^0h_{ij} d\theta_i d\theta_j\big).
\end{equation}
Lowercase Latin letters from the middle of the alphabet denote spatial indices. Here, $N_0$ is the homogeneous lapse function, and $\alpha$ is a dynamical variable that, up to a constant, coincides with the logarithm of the scale factor. Besides, we have chosen spatial angular coordinates $\theta_i$ with a period equal to $2\pi/l_0$, so that $2\pi\theta_i/l_0\in S^1$. We have introduced the (time-independent) Euclidean metric of the three-torus, $^0h_{ij}$, as an auxiliary metric. Finally, we have included the constant factor $\sigma^2= 4\pi/(3l_0^3)$ for convenience. 

In LQG, the fundamental variables that describe the geometry are a densitized triad and an $su(2)$ connection \cite{lqg}, known as the Ashtekar-Barbero variables. Recalling the homogeneity and isotropy of the spatial sections, and making use of an auxiliary triad, e.g. the Euclidean triad $^0e_i$, it is easy to see that each of the Ashtekar-Barbero variables is determined by a single homogeneous variable. 
A usual prescription in LQC \cite{APS2} is to parametrize the densitized triad in terms of a variable, $v$, that is related with the scale factor of the model by \cite{hybr-ref}
\begin{equation}
\label{alfav}
e^{\alpha}=\left( \frac{3\gamma \sqrt{\Delta_g}}{2\sigma}|v|\right)^{1/3}.
\end{equation}
The sign of $v$ determines the orientation of the triad, $\gamma$ is a constant called the Immirzi parameter \cite{immirzi}, and $\Delta_g$ is the minimum nonzero area allowed by the spectrum of the area operator in LQG \cite{APS2}. The variable $|v|$ is proportional to the volume $V$ of the homogeneous sections, which is finite owing to their compactness. We have that $V=2\pi \gamma  \sqrt{\Delta_g}|v|$. In addition, one introduces a canonical variable $b$ that satisfies the Poisson brackets $\{b,v\}=2$. Given the flatness of the spatial sections, it is not difficult to relate $b$ with the time derivative of $\alpha$, and therefore with its standard momentum $\pi_{\alpha}$ in geometrodynamics: 
\begin{equation}
\label{momalfa}
\pi_{\alpha}=-\frac32 v b.
\end{equation}
Expressions \eqref{alfav} and \eqref{momalfa} are the basic formulas to establish the relation between the geometrodynamical and the LQC variables. Since it is much more common to carry out the classical discussion of the system in terms of geometrodynamical variables, we will use them in our analysis up to the point in which one only needs to substitute an operator version of these relations in order to quantize the model according to LQC.  

The matter content of this model will be a scalar field  $\phi(t)$ subject to a potential. For definiteness, we consider the simplest nontrivial case, given by a mass term quadratic in the field. At this stage, we assume that the scalar field is also homogeneous.

Then, this homogeneous model is subject only to one constraint: the Hamiltonian (or scalar) constraint $H_{|0}$, that generates homogeneous time reparametrizations in the system. It can be written \cite{hybr-ref} 
\begin{equation}
\label{H02}
H_{|0} = \frac{e^{-3\alpha}}{2}\left(\frac{3}{4\pi}\pi_\phi^2-\pi_\alpha^2+\frac{4\pi}{3}e^{6\alpha}\tilde{m}^2\phi^2\right)=\frac{3e^{-3\alpha}}{8\pi }\big(\pi_\phi^2- 3\pi v^2b^2+ V^2 m^2 \phi^2\big).
\end{equation}
Here, $m=\tilde{m}/\sigma$ is the mass of the scalar field, and $\pi_{\phi}$ its momentum. 

\subsection{Perturbations of the scalar field and the geometry}

In the model described above, we now perturb the tetrad and the scalar field, and introduce a Dirac field that is regarded as an additional perturbation of the original system. All these perturbations are included up to quadratic contributions in the action, which is the order of truncation in our perturbative approximation \cite{HH,DEH,uf-flat,hybr-ref}. Since the fermionic part of the Einstein-Dirac action is already quadratic in the fermionic contributions, and they are treated as perturbations, including a possible homogeneous component of the Dirac field, the considered truncation is tantamount to couple the Dirac field directly to the tetrad of the homogeneous unperturbed spacetime. Therefore, at the considered perturbative level we only need to study the quadratic perturbation of the Einstein action minimally coupled to a scalar field and add to the result the Dirac action for the fermions evaluated in the FLRW geometry. 

The perturbation of the Einstein action at quadratic order has been studied in a Hamiltonian and manifestly gauge-covariant description in Ref. \cite{hybr-ref}. Since some steps of the treatment presented in that work have implications for our discussion, in this subsection we succinctly review the results that are relevant for our study. The inclusion of fermions will be considered in the next subsection.

Since the perturbations of the metric and the scalar field introduce inhomogeneities, it is convenient to expand them in modes. A general way to introduce well-defined modes and proceed to an expansion of this type is the following. One can consider the connection $^{0}\nabla_i$ of the auxiliary metric $^0h_{ij}$ and its corresponding Laplace-Beltrami operator $ {^0h}^{ij} \,^{0}\nabla_i ^{0}\nabla_j$. The eigenfunctions of this operator provide then a complete set of modes (for functions that are square integrable with the volume element determined by the auxiliary metric). Moreover, these eigenfunctions can be chosen real. We call them $\tilde Q_{\vec n,\epsilon} (\vec\theta)$, with $\vec{\theta}$ being the tuple formed by the coordinates $\theta_i$, and $\epsilon$ a parameter that indicates the behavior under the change of $\theta_i$ by $l_0-\theta_i$: $\epsilon=-1$ if they are odd, and $\epsilon=1$ if they are even. For our flat topology, these eigenfunctions correspond, respectively, to sine and cosine functions. We fix their norm equal to $l_0^{3/2}$ with respect to the auxiliary volume element. The label $\vec{n}=(n_1,n_2,n_3)\in\mathbb Z^3$ is any tuple of integers for which the first nonvanishing component is positive. This label determines the corresponding eigenvalue, which is $-\omega_n^2=-4\pi^2 |n|^2 /l_0^{2}$, with $|n|$ the Euclidean norm of $\vec{n}$. Note that, thanks to the compactness of the spatial sections, the spectrum of the Laplace-Beltrami operator is discrete. Since the zero-modes of the geometry and the scalar field are not considered as perturbations, we obviate these modes from all of our expansions. With our set of eigenfunctions, the connection $^{0}\nabla_i$, and the metric $^0h_{ij}$, we can then construct a complete basis of scalar, vector, and tensor harmonics on the spatial sections.

If we adopt a 3+1 decomposition of the spacetime metric, we can use the above harmonics to expand in modes the spatial metric (a spatial tensor), the shift vector, and the lapse function (a scalar), as well as the scalar matter field. The vector modes do not play any physical role in a system like ours, which does not contain any vector matter field. The genuine tensor perturbations (which are not derived from the scalar harmonics by multiplication by the auxiliary metric nor by differentiation with respect to the associated connection) represent true degrees of freedom of the system. They decouple from the scalar perturbations and their only contribution is the addition of a quadratic term to the zero-mode of the Hamiltonian constraint. They have been studied in detail within the framework of hybrid LQC in Ref. \cite{hybr-ten}. Much more complicated is the treatment of the scalar perturbations. They are also the most relevant perturbations from an observational point of view, since they are responsible of the anisotropies measured in the CMB.

Scalar perturbations affect not only the scalar field, but also (some parts of) the spatial metric, the shift, and the lapse function. Besides, they are constrained by the linearization of the momentum and Hamiltonian constraints of general relativity. The resulting  linear perturbative constraints appear in the quadratic truncation of the perturbed action (written in Hamiltonian form) accompanied by Lagrange multipliers that are determined by the perturbed lapse and shift \cite{HH,hybr-ref}. The need to consider quantities that are invariant under the transformations generated by these perturbative constraints, so that they have a well-defined physical meaning, leads to the introduction of the so-called gauge invariants \cite{bardeen,gaugeinvariant}. In particular, it is especially convenient to use the Mukhanov-Sasaki invariant field \cite{sasa,kodasasa,mukhanov,mukh}, since it is directly related with the comoving curvature perturbations, and therefore with the power spectrum of the CMB anisotropies. This gauge invariant is obtained from an appropriate (nonlocal) combination of the scalar matter field perturbation and the scalar perturbations of the metric. In our analysis, we will describe it by the (time-dependent) coefficients $v_{\vec{n},\epsilon}$ of its mode expansion in the basis formed by $\tilde Q_{\vec n,\epsilon} (\vec\theta)$.

It was proven in Ref. \cite{hybr-ref} that, at the level of our perturbative truncation, it is possible to Abelianize the linear perturbative constraints and, with their Abelian version and the Mukhanov-Sasaki gauge invariant, form a complete set of compatible perturbative variables, in the sense that they commute under Poisson brackets. This set can be completed into a canonical one introducing suitable momenta, in such a way that the momentum of the Mukhanov-Sasaki field is also a gauge invariant. The freedom in the choice of this momentum is removed by demanding that, dynamically, it is proportional to the time derivative of the Mukhanov-Sasaki field on shell. We will call $\pi_{v_{\vec{n},\epsilon}}$ the (time-dependent) coefficients of the mode expansion of this momentum. 

Much more remarkably, the analysis of Ref. \cite{hybr-ref} demonstrated also that the variables of the homogeneous system can be corrected with quadratic terms in the perturbations so as to complete the canonical set of perturbative variables into a canonical set for the entire system (homogeneous sector plus perturbations), at the level of our perturbative approximation. The explicit expression of these new homogeneous variables can be found in Appendix A of Ref. \cite{hybr-ref}. The modification can be regarded as a type of backreaction effect of the perturbations onto the definition of the homogeneous variables. 

Apart from the linear perturbative constraints, only one more constraint remains to be imposed, which in fact is a global one. It is the zero-mode of the Hamiltonian constraint. As far as the geometry and the matter scalar field are concerned, it is the sum of the Hamiltonian constraint $H_{|0}$ of the homogeneous system (evaluated in the corrected, new homogeneous variables) plus a quadratic contribution of the Mukhanov-Sasaki invariant and its momentum, that we call the Mukhanov-Sasaki Hamiltonian and denote by ${\breve{H}}_{|2}$ (with the notation of Ref.\cite{hybr-ref}). In addition, if there are tensor perturbations, the constraint includes also a quadratic contribution of them, that we call ${^T{\tilde{H}}_{|2}}$ \cite{hybr-ten}. Using from now on the notation $(\alpha,\pi_{\alpha},\phi,\pi_{\phi})$ [and its LQC counterpart $(v,b,\phi,\pi_{\phi})$] to denote the perturbatively corrected homogeneous variables instead of the original ones, the quadratic perturbative contributions to the Hamiltonian constraint can be expressed as \cite{hybr-ref,hybr-ten}
\begin{align}
\label{H2} 
{\breve{H}}_{|2}&=\sum_{\vec n,\epsilon}\breve H^{\vec n,\epsilon}_{|2},\qquad
{^T{\tilde{H}}_{|2}}= \sum_{\vec n,\epsilon,\tilde{\epsilon}} {^{T}{\tilde H}^{\vec {n},\epsilon,\tilde{\epsilon}}_{|2}},\end{align}
\begin{eqnarray}
\label{H2MS}
\breve H^{\vec n,\epsilon}_{|2}&=&\frac{e^{-{\alpha}}}{2}\bigg[ \omega_n^2 + e^{-4{\alpha}}\pi_{\alpha}^2+\tilde{m}^2 e^{2{\alpha}}\left(1+20\pi{\phi}^2-12{\phi} \frac{\pi_{\phi}}{\pi_{\alpha}}-32\pi^2 e^{6{\alpha}}\tilde{m}^2 \frac{{\phi}^4}{\pi_{\alpha}^2} \right)\bigg] v_{\vec{n},\epsilon}^2 \nonumber\\
&+&\frac{e^{-{\alpha}}}{2} \pi_{v_{\vec{n},\epsilon}}^2,\\
\label{TH_2}
{^{T}{\tilde H}^{\vec {n},\epsilon,\tilde{\epsilon}}_{|2}} &=& \frac{e^{-{\alpha}}}{2}\big[\big(\omega_n^2+e^{-4 \alpha}\pi_\alpha^2
-4\pi{\tilde{m}}^2 e^{2\alpha}\phi^2\big){\tilde d}_{\vec {n},\epsilon,\tilde{\epsilon}}^2+\pi_{\tilde d_{\vec {n},\epsilon,\tilde{\epsilon}}}^2\big].
\end{eqnarray}
Here, $\tilde{\epsilon}$ is a dichotomous label that describes the two possible polarizations of the tensor perturbations ($\tilde{\epsilon}=+,\times$), the variables ${\tilde d}_{\vec {n},\epsilon,\tilde{\epsilon}}$ are the (time-dependent) coefficients of the mode expansion of the tensor perturbations multiplied by $e^{\alpha}$ (i.e., the scale factor up to a constant), and $\pi_{\tilde d_{\vec {n},\epsilon,\tilde{\epsilon}}}$ are their canonical momenta \cite{hybr-ten}. The freedom to add a contribution linear in ${\tilde d}_{\vec {n},\epsilon,\tilde{\epsilon}}$ to these momenta has been fixed by choosing them so that, dynamically, they are proportional to the time derivatives of their configuration variables. 

\subsection{Dirac fermions as perturbations}

We now proceed to include a Dirac field in the system, treated as a perturbation of the homogeneous model. As we have already commented, this field is governed by the fermionic part of the Einstein-Dirac action, which is quadratic in the fermions, and therefore in the perturbations \cite{DEH}. Hence, at our order of perturbative truncation, the coupling with the tetrad in the action for the Dirac field can be replaced with a coupling with the tetrad of the homogeneous spacetime, either before or after correcting it with the modifications introduced by the scalar perturbations. The difference between those tetrads, multiplied by a quadratic term in the fermions, is of higher than second order in the perturbations, and hence can be neglected in our approximations. Thus, in the following, we consider that the Dirac fermion is directly coupled to the homogeneous geometry with the corrected scale factor. As we explained at the end of the previous subsection, to avoid complicating in excess our notation, we maintain our original symbols for the homogeneous variables in spite of having already modified them with quadratic terms of the scalar perturbations, as worked out in Ref. \cite{hybr-ref}.

It is clear that the homogeneous and isotropic spacetime with scale factor $a=\sigma e^{\alpha}$ admits a global orthonormal tetrad, that we call $e^{\mu}_{a}$, where, $\mu=0,1,2,3$ is a spacetime index and $a=0,1,2,3$ is an internal gauge index. As a consequence, we can always define a spin structure \cite{Geroch,SGeom}. We will keep our discussion general and do not make explicit the choice of this structure. 
The Dirac field $\Psi$ can be understood as the cross sections of the corresponding spinor bundle, such that they obey the Dirac equation, with a mass $M$,
\begin{equation}\label{Deq}
e^{\mu}_{a}\gamma^{a}\nabla^{S}_{\mu}\Psi=M\Psi.
\end{equation}
In turn, the fermionic part of the Einstein-Dirac action has the form
\begin{equation}
\label{Dirac4action}
I_D= \int d{\cal V} \left[ i M  \Psi^{\dagger}\gamma^0 \Psi -\frac{1}{2}\Big(i \Psi^{\dagger} \gamma^0 e^{\mu}_a \gamma^a \nabla_{\mu}^{S}\Psi + {\mathrm{Hermitian\; conjugate}}\Big) \right].
\end{equation}
Here, $d{\cal  V}$ is the four-dimensional volume element corresponding to the considered spacetime metric, the dagger denotes Hermitian conjugate, $\nabla^{S}_{\mu}$ is the spin lifting of the Levi-Civit\`{a} connection \cite{SGeom}, and $\gamma^{a}$ are the Dirac matrices. Adopting the so-called Weyl representation for them, as in Ref. \cite{DEH}, we can describe the Dirac field by a pair of two-component spinors of definite chirality. The left-handed and right-handed projections of $\Psi$ will be called $\phi^{A}$ and $\bar{\chi}_{A'}$, respectively, where $A=1,2$, $A'=1',2'$, and the bar denotes complex conjugation. The components of these spinors are Grassman variables \cite{Berezin}, to reflect the anticommuting properties of fermions. Besides, spinor indices are raised and lowered with the antisymmetric matrices $\epsilon^{AB}$, $\epsilon_{AB}$, $\epsilon^{A'B'}$, and $\epsilon_{A'B'}$, all of them with nondiagonal component equal to one if the chiral indices that label them are in increasing order \cite{DEH}.

Following the treatment of Ref. \cite{DEH}, it is convenient to introduce a choice of gauge fixing known as time gauge, imposing that  $e^{j}_{0}=0$. Notice that this fixes only part of the internal gauge, intrinsic to the fermions, but does not affect at all the gravitational constraints (perturbative or not) of the rest of the system. The effect on the spin structure is a restriction that can be reinterpreted as a spin structure on each of the toroidal spatial sections. The two component spinors of the Dirac field can be seen as families of cross sections of the resulting spinor bundle, parametrized by the time coordinate \cite{uf-flat}. On the other hand, the anticommutation canonical relations of the Dirac field are provided by the symmetric Dirac brackets at coincident time
\begin{equation}\label{antibracket}
\{a^{3/2}\Psi^{\dagger}(\vec{\theta}),a^{3/2} \Psi(\vec{\theta}')\}=-i\delta(\vec{\theta}-\vec{\theta}')I,
\end{equation}
where $I$ is the identity matrix in four dimensions. These brackets are obtained after eliminating second-class constraints that relate the Dirac field with its momentum, and that appear in the system because the Dirac action is first-order in the field derivatives.

Similar to our analysis of the scalar and tensor perturbations, we can decompose the two-component spinors of the Dirac field in modes. In the present case, it is convenient to choose the basis of spinor modes as a basis of eigenfunctions of the Dirac operator associated with the auxiliary Euclidean triad on the toroidal sections. The corresponding spectrum is discrete, with eigenvalues $\pm \omega_k$ given by $\omega_{k}=2\pi|\vec{k}+\vec{\tau}|/l_0$ \cite{Dtorus}.
Vertical bars symbolize the Euclidean norm, and the row vector $\vec{k}\in\mathbb{Z}^{3}$ is any tuple of integers. We have used the notation $\vec{\tau}=\sum_I \tau^I \vec{v}_I/2$, where $I=1,2,3$, the vectors $\vec{v}_I$ form the standard orthonormal basis of the lattice $\mathbb{Z}^{3}$, and $\tau^{I}\in\{0,1\}$ characterizes each of the possible spin structures on $T^{3}$. The label $k$ in $\omega_{k}$ distinguishes eigenvalues, and can be identified with the norm of one of the tuples $\vec{k}$ that leads to it. The Dirac eigenvalues can be degenerated, with degeneracy denoted by $g_k$. This degeneracy grows as a function of asymptotic order ${\mathcal{O}}(\omega_k^2)$ when $k$ tends to infinity.

With the choice of the Euclidean triad as the auxiliary one, the spin connection vanishes. One can then obtain the explicit form of the Dirac eigenspinors. For the left-handed chirality of $\phi_A$ and the eigenvalues $\pm\omega_{k}$, one gets
\begin{equation}\label{eigens}
w^{\vec{k},(\pm)}_{A}=u^{\vec{k},(\pm)}_{A}e^{i \frac{2\pi}{l_0} ( \vec{k}+\vec{\tau})\cdot\vec{\theta}}, 
\end{equation}
where the constant two-component spinors $u^{\vec{k},(\pm)}_{A}$ are normalized so that
\begin{equation}
\bar{u}^{\vec{k},(\pm)}_{1'} u^{\vec{k},(\pm)}_1+\bar{u}^{\vec{k},(\pm)}_{2'} u^{\vec{k},(\pm)}_2=1
\end{equation}
and, for all $\vec{k},\vec{k}'\neq -\vec{\tau}$, they satisfy the conditions
\begin{equation}\label{norm}
 u^{\vec{k}',(+)}_{A}\epsilon^{AB}u^{\vec{k},(-)}_{B}=0,\qquad u^{\vec{k}',(\pm)}_{A}\epsilon^{AB}u^{\vec{k},(\pm)}_{B}=e^{iC^{(\pm)}_{\vec{k}}}\delta_{\vec{k}',-\vec{k}-2\vec{\tau}} \,.
\end{equation}
Summation over repeated spinorial indices is assumed, and $C^{(\pm)}_{\vec{k}}$ are constants that can be changed by modifying the phase of $u^{\vec{k},(\pm)}$. The above conditions do not apply in the case of zero-modes, namely, for modes with vanishing eigenvalue $\omega_k$. These modes exist only when the spin structure is trivial ($\vec{\tau}=0$). For such modes, one can take $u^{\vec{0},(+)}_A$ as the spinor defined by $u^{\vec{0},(+)}_1=1$ and $u^{\vec{0},(+)}_2=0$, whereas for $u^{\vec{0},(-)}_A$ one takes $u^{\vec{0},(-)}_1=0$ and $u^{\vec{0},(-)}_2=1$. On the other hand, the complex conjugate of Eq. \eqref{eigens} provides a basis of modes for eigenspinors of right-handed chirality, like $\bar{\chi}_{A'}$.

Let us expand, in the above basis of eigenmodes of the Dirac operator, our field multiplied by $a^{3/2}=\sigma^{3/2}e^{3\alpha/2}$ [as it appears in the anticommutation relations \eqref{antibracket}] and by a convenient constant factor $l_0^{3/2}$ (equal to the square root of the auxiliary volume of the toroidal sections). Let us call $m_{\vec{k}}$ and $\bar{r}_{\vec{k}}$ the time-dependent coefficients in the mode expansion of the left-handed spinor for positive and negative eigenvalues, respectively, and $\bar{s}_{\vec{k}}$ and $t_{\vec{k}}$ the corresponding coefficients for the right-handed part:
\begin{eqnarray} \label{modeexpansion}
\phi_A(x)&=&\frac{e^{-3\alpha/2}}{\sigma^{3/2}l_0^{3/2}}\sum_{\vec{k},(\pm)}\big[m_{\vec{k}}w^{\vec{k},(+)}_{A}+\bar{r}_{\vec{k}}w^{\vec{k},(-)}_{A}\big],\\
\bar{\chi}_{A'}(x)&=&\frac{e^{-3\alpha/2}}{\sigma^{3/2}l_0^{3/2}}\sum_{\vec{k},(\pm)}\big[\bar{s}_{\vec{k}}\bar{w}^{\vec{k},(+)}_{A'}+t_{\vec{k}}\bar{w}^{\vec{k},(-)}_{A'}\big].
\end{eqnarray}
Let us also denote generically as $(x_{\vec{k}},y_{\vec{k}})$ any of the ordered pairs $(m_{\vec{k}},s_{\vec{k}})$ or $(t_{\vec{k}},r_{\vec{k}})$. Then, the Dirac equation \eqref{Deq} leads to the following set of dynamical equations for all $\vec{k}\neq \vec{\tau}$ \cite{uf-flat}:  
\begin{align}\label{1order}
x_{\vec{k}}'=i\omega_{k}x_{\vec{k}}-i\tilde{M}e^{\alpha}\bar{y}_{-\vec{k}-2\vec{\tau}}\;, \qquad \bar{y}_{\vec{k}}'=-i\omega_{k}\bar{y}_{\vec{k}}-i\tilde{M}e^{\alpha}x_{-\vec{k}-2\vec{\tau}}\;,
\end{align}
where the prime stands for the derivative with respect to a conformal time $\eta$ defined via $d\eta=e^{-\alpha}dt$, and $\tilde{M}=M\sigma$. 

Introducing the representation of the Dirac field in terms of two-component spinors and the mode expansion of the latter, one then arrives to a expression of the action in terms of mode coefficients \cite{DEH}:
\begin{equation}
\label{Diracmodeaction}
I_D=\delta_{\vec{0}}^{\vec\tau} I_{\vec{0}} + \sum_{\vec{k}\neq\vec\tau} I_{\vec{k}},
\end{equation}
where $\delta_{\vec{0}}^{\vec\tau}=1$ if $\vec\tau=\vec{0}$ and vanishes otherwise. Here, the contribution of the nonzero-modes is
\begin{eqnarray}
\label{Diracmodecontribution}
I_{\vec{k}}&=& \int dt \left[ -\frac{i}{2} \big(\dot{m}_{\vec{k}}\bar{m}_{\vec{k}}+\dot{\bar{m}}_{\vec{k}}m_{\vec{k}}
+\dot{r}_{\vec{k}}\bar{r}_{\vec{k}}+\dot{\bar{r}}_{\vec{k}}r_{\vec{k}}
+\dot{s}_{\vec{k}}\bar{s}_{\vec{k}}+\dot{\bar{s}}_{\vec{k}}s_{\vec{k}}
+\dot{t}_{\vec{k}}\bar{t}_{\vec{k}}+\dot{\bar{t}}_{\vec{k}}t_{\vec{k}}\big) \right.\nonumber \\
&-& N_0 \tilde{M}  \big( {s}_{-\vec{k}-2\vec\tau} m_{\vec{k}} + {\bar{m}}_{\vec{k}} {\bar s}_{-\vec{k}-2\vec\tau} + r_{-\vec{k}-2\vec\tau} t_{\vec{k}} + {\bar{t}}_{\vec{k}} {\bar r}_{-\vec{k}-2\vec\tau} \big) \nonumber\\
&+& N_0 e^{-\alpha} \omega_k \left. \big( {\bar{m}}_{\vec{k}} m_{\vec{k}} + {\bar{t}}_{\vec{k}} t_{\vec{k}} - r_{\vec{k}} {\bar r}_{\vec{k}} - s_{\vec{k}} {\bar s}_{\vec{k}} \big) \right],
\end{eqnarray}
where the dot means derivative with respect to $t$. For the trivial spin structure, one has to add the zero-mode contribution
\begin{eqnarray}
\label{Diraczeromode}
I_{\vec{0}}&=& \int dt \left[ -\frac{i}{2} \big(\dot{m}_{\vec{0}}\bar{m}_{\vec{0}}+\dot{\bar{m}}_{\vec{0}}m_{\vec{0}}
+\dot{r}_{\vec{0}}\bar{r}_{\vec{0}}+\dot{\bar{r}}_{\vec{0}}r_{\vec{0}}
+\dot{s}_{\vec{0}}\bar{s}_{\vec{0}}+\dot{\bar{s}}_{\vec{0}}s_{\vec{0}}
+\dot{t}_{\vec{0}}\bar{t}_{\vec{0}}+\dot{\bar{t}}_{\vec{0}}t_{\vec{0}}\big) \right.\nonumber \\
&-& N_0 \tilde{M}  \left. \big( {s}_{\vec{0}} {\bar r}_{\vec{0}} + r_{\vec{0}} {\bar s}_{\vec{0}} + m_{\vec{0}} {\bar t}_{\vec{0}} + t_{\vec{0}} {\bar m}_{\vec{0}} \big)  \right].
\end{eqnarray}

The part with time derivatives determines the anticommutation relations in terms of the mode coefficients, and tells us that the pairs $(x_{\vec{k}},\bar{y}_{\vec{k}})$ are canonical Grassman variables. The rest of the action, that is linear in the homogeneous lapse function, supplies a contribution to the zero-mode of the Hamiltonian constraint of the total system, contribution that, as we expected, is quadratic in the fermionic variables. Namely, the global Hamiltonian constraint of our perturbed system becomes
\begin{eqnarray}
\label{HTotalF}
H_{|0}&+&{\breve{H}}_{|2}+ {^{T}{\tilde{H}}_{|2}}+H_D,\qquad H_D=\delta_{\vec{0}}^{\vec{\tau}} H_{\vec{0}} + \sum_{\vec{k}\neq\vec{\tau}} H_{\vec{k}},\\
\label{HkF}
H_{\vec{k}}&=& 
\tilde{M}  \big( {s}_{-\vec{k}-2\vec\tau} m_{\vec{k}} + {\bar{m}}_{\vec{k}} {\bar s}_{-\vec{k}-2\vec\tau} + r_{\vec{k}} t_{-\vec{k}-2\vec\tau} + {\bar{t}}_{-\vec{k}-2\vec\tau} {\bar r}_{\vec{k}} \big) \nonumber\\
&-&  e^{-\alpha} \omega_k  \big( {\bar{m}}_{\vec{k}} m_{\vec{k}} + {\bar{t}}_{\vec{k}} t_{\vec{k}} - r_{\vec{k}} {\bar r}_{\vec{k}} - s_{\vec{k}} {\bar s}_{\vec{k}} \big),\\
\label{H0F}
H_{\vec{0}}&=& \tilde{M}  \big( {s}_{\vec{0}} {\bar r}_{\vec{0}} + r_{\vec{0}} {\bar s}_{\vec{0}} + m_{\vec{0}} {\bar t}_{\vec{0}} + t_{\vec{0}} {\bar m}_{\vec{0}} \big).
\end{eqnarray}

Finally, we notice that (after having corrected the homogeneous variables with suitable quadratic contributions of the scalar perturbations) the system is symplectic at our order of perturbative truncation, and it is described by a homogeneous part, the Mukhanov-Sasaki  gauge invariant and its momentum, the mode coefficients of the tensor perturbations (rescaled with the scale factor) and their momenta, and the above fermion mode coefficients, all this in addition to the linear perturbative constraints and some suitable momenta of them (which form canonical pairs that commute with all the previous variables).  Apart from those linear perturbative constraints, our system is subject only to the zero-mode of the Hamiltonian constraint specified above, and to gauge rotations of the fermions (the rest of internal gauge transformations have been fixed when we have imposed the time gauge \cite{DEH}). Since the role of gauge transformations is well under control and is not crucial for the passage to the quantum theory, we simply assume that the remaining gauge freedom has also been fixed, e.g. by choosing a certain triad among all those related by gauge rotations. This leaves the constraint \eqref{HTotalF} as the only remaining one. 

\section{Creation and annihilation variables for the Dirac field}

Before we can proceed to the hybrid quantization of our system, in which we will adopt a Fock representation for the fermionic variables, we have to introduce creation and annihilation variables for the Dirac field. There is an infinite ambiguity in their definition that, in our model, reflects the freedom of choice at two stages. On the one hand, one can extract part of the evolution of the fermionic variables and express it in terms of the background, role which is played in our system by the homogeneous sector. If one considers only the dynamical effect of the background geometry, the available redefinitions of the creation and annihilation variables and of their evolution are those related by transformations that depend on the (homogeneous) scale factor and its momentum. On the other hand, even if we select the dynamics, we can still choose different sets of creation and annihilation variables that could lead to inequivalent Fock representations. The different possibilities correspond to choices of different complex structures \cite{wald}. These sets of variables are now related by constant transformations, since the dynamical content has already been taken into account in the first step of our considerations. Obviously, the combined freedom of choice is described by all possible $(\alpha,\pi_{\alpha})$-dependent canonical transformations. Given the linearity of the field equations and of the basic structures for a Fock representation, like the complex structure, we will restrict our attention to linear transformations. Besides, we ask them to respect the dynamical decoupling between modes, although they may be mode dependent: they may vary with the labels that characterize each Dirac mode. In total, we analyze creation and annihilation variables of the generic form:
\begin{eqnarray}
\label{avariable}
a_{\vec{k}}^{(x,y)}&=& f_1^{\vec{k},(x,y)}(\alpha,\pi_{\alpha})\, x_{\vec{k}}+f_2^{\vec{k},(x,y)}(\alpha,\pi_{\alpha})\, {\bar y}_{-\vec{k}-2\vec{\tau}}, \nonumber\\
{\bar b}_{\vec{k}}^{(x,y)}&=& g_1^{\vec{k},(x,y)}(\alpha,\pi_{\alpha})\, x_{\vec{k}}+g_2^{\vec{k},(x,y)}(\alpha,\pi_{\alpha}) \,{\bar y}_{-\vec{k}-2\vec{\tau}},	
\end{eqnarray}
where we recall that $\vec{\tau}$ is fixed and differs for each of the allowed spin structures. The background-dependent coefficients in these linear expressions of the fermionic variables may change for positive and negative helicity, corresponding to the pair $(x_{\vec{k}},y_{\vec{k}})=(m_{\vec{k}},s_{\vec{k}})$ or to $(x_{\vec{k}},y_{\vec{k}})=(t_{\vec{k}},r_{\vec{k}})$. In a Fock representation with a standard interpretation, the operators for $a_{\vec{k}}^{(x,y)}$ and ${\bar b}_{\vec{k}}^{(x,y)}$ would annihilate particles and create antiparticles, respectively. 

Based on recent results about criteria to remove the ambiguity in the choice of a Fock quantization on cosmological backgrounds, first proposed for scalar fields \cite{unique1,unique2,unique3,unique4} and then extended to fermions \cite{uf1,uf2,uf3}, and in particular to Dirac fields in flat FLRW spacetimes \cite{uf-flat}, we can minimize the physical consequences of our freedom of choice of creation and annihilation variables. Under the requirements of: i) unitary implementability of the dynamics of the chosen variables in the quantum theory, ii) invariance of this theory under the Killing isometries of the toroidal sections and the spin rotations generated by the helicity, and iii) a convention for the concepts of particles and antiparticles that connects smoothly in the massless limit with the standard one, the analysis of Ref. \cite{uf-flat} demonstrated that the family of possible choices of variables has associated Fock representations which are all unitarily equivalent. This family is precisely of the form \eqref{avariable}, with background-dependent coefficients restricted by our three requirements (of unitarity, invariance, and a standard convention of particles and antiparticles) as follows, except perhaps for a finite number of modes:
\begin{itemize}
		\item[a)] For tuples $\vec{k}$ in an infinite subset $\mathbb{Z}_{1}^{3}$ of $\mathbb{Z}^{3}$, the functions $f_{1}^{\vec{k},(x,y)}$ have the asymptotic behavior at large $|\vec{k}|$ and at all times:
		\begin{equation}\label{unith}
		f^{\vec{k},(x,y)}_{1}=\frac{\tilde{M}e^{\alpha}}{2\omega_{k}}e^{iF^{\vec{k},(x,y)}_{2}}+\vartheta^{\vec{k},(x,y)}\quad {\mathrm{with}}\quad
		\sum_{\vec{k}} \left\vert\vartheta^{\vec{k},(x,y)}\right\vert^{2}<\infty .
		\end{equation}	
		\item[b)] If the complement of $\mathbb{Z}_{1}^{3}$ in $\mathbb{Z}^{3}$ is infinite, for tuples $\vec{k}$ belonging to it the functions $f_{1}^{\vec{k},(x,y)}$ must be asymptotically of order $\omega_{k}^{-1}$ or higher, and form a sequence that is square summable at all times.
\end{itemize}
Besides, the rest of coefficients must satisfy the relations
\begin{eqnarray}
\label{g1,g2}
g^{\vec{k},(x,y)}_{1} &=& e^{iG^{\vec{k},(x,y)}}{\bar f}^{\vec{k},(x,y)}_{2},\qquad g^{\vec{k},(x,y)}_{2} = - e^{iG^{\vec{k},(x,y)}} {\bar f}^{\vec{k},(x,y)}_{1} ,\\
\label{f2}
f^{\vec{k},(x,y)}_{2}&=& e^{iF^{\vec{k},(x,y)}_{2}} \sqrt{1- \left\vert f^{\vec{k},(x,y)}_{1} \right\vert^2}.
\end{eqnarray}
To obtain these results, Ref. \cite{uf-flat} used some mild assumptions about the logarithmic scale factor $\alpha$, e.g. that it has a continuous third derivative with respect to the conformal time. 

In the rest of this section, we will study the consequences of adopting a set of creation and annihilation variables with the above properties. Later on, in Sec. IV, we will particularize our analysis to a specific set of this type, namely, the variables employed in Ref. \cite{DEH} by D'Eath and Halliwell, adapted to our case of toroidal spatial sections.

Since our change of fermionic variables from the pairs $(x_{\vec{k}},y_{\vec{k}})$ to the pairs $(a_{\vec{k}}^{(x,y)},b_{\vec{k}}^{(x,y)})$ may depend on the homogeneous logarithmic scale factor $\alpha$ and its momentum $\pi_{\alpha}$, in general they do not longer commute with this homogeneous pair. To recover the canonical structure of our system, we must modify the homogeneous variables, correcting them with fermionic contributions that counterbalance the loss of commutativity. The calculation can be carried out in a way similar to that presented in Sec. 4.1 of Ref. \cite{hybr-ref}, but now extended to the consideration of Grasmman variables. One starts with the Legendre term of the perturbed action of our system (truncated at second perturbative order), symmetrized in the fermionic $(x,y)$-variables, and introduces the inverse of the linear change \eqref{avariable}. Integrating by parts time derivatives in the fermionic contribution, disregarding irrelevant boundary terms (evaluated at initial and final times), and neglecting terms of higher than second order in the fermions, it is not difficult to see that the corrected homogeneous variables, that render the system canonical again, are given by
\begin{eqnarray}
\label{modifiedscale}
\breve{\alpha}&=&\alpha +\frac{i}{2}\sum_{\vec{k},(x,y)}[
(\partial_{\pi_{\alpha}}x_{\vec{k}})  {\bar x}_{\vec{k}}+(\partial_{\pi_{\alpha}}{\bar x}_{\vec{k}})  x_{\vec{k}}+(\partial_{\pi_{\alpha}}y_{\vec{k}})  {\bar y}_{\vec{k}}+(\partial_{\pi_{\alpha}}{\bar y}_{\vec{k}})  y_{\vec{k}} ],
\\
\label{modifiedpi}
{\breve{\pi}}_{\alpha}&=&\pi_{\alpha}-\frac{i}{2}\sum_{\vec{k},(x,y)}[
(\partial_{\alpha}x_{\vec{k}})  {\bar x}_{\vec{k}}+(\partial_{\alpha}{\bar x}_{\vec{k}})  x_{\vec{k}}+(\partial_{\alpha}y_{\vec{k}})  {\bar y}_{\vec{k}}+(\partial_{\alpha}{\bar y}_{\vec{k}})  y_{\vec{k}} ],
\end{eqnarray} 
where the sum over $(x,y)$ is over left-handed and right-handed chiral pairs, $(m,s)$ and $(t,r)$.

A especially interesting situation, given its simplicity, is the case in which the phases $G^{\vec{k},(x,y)}$ and $F^{\vec{k},(x,y)}_{2}$ of the coefficients \eqref{g1,g2} and \eqref{f2}, as well as the phase of $f^{\vec{k},(x,y)}_{1}$, are constant. Then, a straightforward calculation, using the inverse of the linear relation \eqref{avariable} and Eqs. \eqref{g1,g2} and \eqref{f2}, shows that the fermionic corrections in the definition of the canonical pair of variables for the homogeneous geometry are a linear combination of the products $a_{\vec{k}}^{(x,y)}b_{\vec{k}}^{(x,y)}$ and ${\bar b}_{\vec{k}}^{(x,y)}{\bar a}_{\vec{k}}^{(x,y)}$, with coefficients that are complex conjugates one of each other.

Returning to the general case, we notice that, in terms that are exactly quadratic in the perturbations, the replacement of our homogeneous variables for the geometry with the new ones has no effect at our order of truncation, since the difference between the two considered sets of homogeneous variables is also quadratic, and would produce new terms that are at least quartic in the perturbations. Therefore, for all such terms, we can simple substitute the pair $(\alpha,\pi_{\alpha})$ by $(\breve{\alpha},{\breve \pi}_{\alpha})$. This is not the case, however, for homogeneous contributions that depend on the geometry. In our Hamiltonian description of the system, the only term of this type is the homogeneous part of the zero-mode of the Hamiltonian constraint, $H_{|0}$. Following the same procedure as in Sec. 4.2 of Ref. \cite{hybr-ref}, if we insert the expression of the old homogeneous variables in terms of the new ones in the geometric dependence of $H_{|0}$, expand the result around the new homogeneous pair, and truncate it at second perturbative order, consistently with the rest of our approximations, we obtain in place of $H_{|0}(\alpha,\pi_{\alpha})$ the term
\begin{equation}
\label{H0new}
H_{|0}(\breve{\alpha},{\breve \pi}_{\alpha})-\Delta{\breve \alpha}\, \partial_{\alpha}H_{|0}(\breve{\alpha},{\breve \pi}_{\alpha})-\Delta{\breve \pi}_{\alpha} \partial_{\pi_{\alpha}}H_{|0}(\breve{\alpha},{\breve \pi}_{\alpha}),
\end{equation}
where $H_{|0}(\breve{\alpha},{\breve \pi}_{\alpha})$ is the original homogeneous Hamiltonian with the old geometric variables \emph{identified} with the new ones. Besides, $\Delta{\breve \alpha}=\breve{\alpha}-\alpha$ and $\Delta{\breve \pi}_{\alpha}=\breve{\pi}_{\alpha}-\pi_{\alpha}$, both quantities expressed in terms of the new variables for the homogeneous geometry and of the fermionic creation and annihilation variables. Given Eqs. \eqref{modifiedscale} and \eqref{modifiedpi}, these last two quantities are quadratic in the fermions, and hence in the perturbations. Using the explicit expression \eqref{H02} of $H_{|0}$, we can rewrite Eq. \eqref{H0new} as
\begin{equation}
\label{H0new1}
H_{|0}(\breve{\alpha},{\breve \pi}_{\alpha})+\left[3 H_{|0}(\breve{\alpha},{\breve \pi}_{\alpha})-4\pi e^{3\breve\alpha}\tilde m^2\phi^2\right]\Delta\breve\alpha+e^{-3\breve{\alpha}}{\breve \pi}_{\alpha}\Delta\breve\pi_{\alpha}.
\end{equation}
In this way, and up to a contribution that is a sum of the linear perturbative constraints (with coefficients that are also linear in the perturbations \cite{hybr-ref}), we get that the total Hamiltonian of the system is
\begin{equation}
\label{operating}
N_0\left(H_{|0}+3 H_{|0} \Delta\breve\alpha-4\pi e^{3\breve\alpha}\tilde m^2\phi^2\Delta\breve\alpha+e^{-3{\breve \alpha}}{\breve \pi}_{\alpha}\Delta\breve\pi_{\alpha} +\breve{H}_{|2}+ {^T \tilde{H}_{|2}} + H_D[a,b]\right),
\end{equation}
where all the dependence on the homogeneous geometry must be evaluated setting the original variables $(\alpha,\pi_{\alpha})$ equal to $(\breve{\alpha},{\breve \pi}_{\alpha})$, and we have called $H_D[a,b]$ the fermionic Hamiltonian $H_D$ expressed in terms of our creation and annihilation variables. 

Since the second term in the above formula is proportional to $H_{|0}$, with a proportionality factor that is quadratic in the fermions, we can absorb it at our order of perturbative truncation by redefining the lapse as $\breve{N}_0=N_0+3\Delta\breve\alpha$. On the other hand, the quadratic contribution of the fermions in the Hamiltonian constraint is given by 
\begin{equation}
\label{breveHD}
{\breve H}_D= -4\pi e^{3\breve\alpha}\tilde m^2\phi^2\Delta\breve\alpha+e^{-3{\breve \alpha}}{\breve \pi}_{\alpha}\Delta\breve\pi_{\alpha} + H_D[a,b].
\end{equation}
The first and second of these terms are just the change in the fermionic Hamiltonian owing to the fact that our change to creation and annihilation variables is time dependent via its dependence on the homogeneous geometry. Summarizing, at our truncation order and modulo the linear perturbative constraints, we finally obtain the total Hamiltonian 
\begin{equation}
\label{breveHami}
{\breve N}_0\big(H_0 +\breve{H}_{|2}+ {^T \tilde{H}_{|2}} + {\breve H}_D\big).
\end{equation}

\section{Variables for instantaneous diagonalization}

Let us now particularize our discussion to a set of creation and annihilation variables similar to that used for the description of the Dirac field by D'Eath and Halliwell \cite{DEH}. These variables have the distinctive property of allowing an instantaneous diagonalization of the Hamiltonian $H_D$ (ignoring zero-modes). They are determined by the choice of coefficients
\begin{eqnarray}
\label{DEHfs}
f^{\vec{k},(x,y)}_{1}&=&\sqrt{ \frac{\xi_k -\omega_k}{2\xi_k} },\qquad 
f^{\vec{k},(x,y)}_{2}=\sqrt{ \frac{\xi_k +\omega_k}{2\xi_k}}, \\
g^{\vec{k},(x,y)}_{1} &=& f^{\vec{k},(x,y)}_{2},\qquad g^{\vec{k},(x,y)}_{2} = - f^{\vec{k},(x,y)}_{1} ,
\end{eqnarray}
where 
\begin{equation}
\label{xik}
\xi_k=\sqrt{\omega_k^2+\tilde{M}^2 e^{2\alpha}}.
\end{equation}
We note that $\xi_k\geq \omega_k>0$. Hence, the $f$-coefficients in Eq. \eqref{DEHfs} are well-defined and real, and the expression of the $g$-coefficients, that are real as well, coincides with the particularization of formula \eqref{g1,g2} to our case. Moreover, it is not difficult to check that the above set of coefficients possesses the asymptotic behavior \eqref{unith}, so that the corresponding variables lead to a Fock representation in the privileged family selected by our uniqueness criterion of unitary dynamics, symmetry invariance, and standard convention of particles and antiparticles. As we have commented, this choice of variables turns out to diagonalize the part of the nonzero-modes in the Hamiltonian $H_D$. Recall that this is not our full Hamiltonian ${\breve H}_D$. In this sense, it is worth noticing that, even if the issue of particle production was discussed in Ref. \cite{DEH} in terms of the creation and annihilation variables defined by Eqs. \eqref{DEHfs}, the Hamiltonian treatment was carried out using in fact the variables $\{m_{\vec{k}},r_{\vec{k}},s_{\vec{k}},t_{\vec{k}}\}$ for the fermionic nonzero-modes, without modifying $H_D$, and introducing a holomorphic representation in the passage to the quantum theory rather than a Fock one. 

Since the coefficients \eqref{DEHfs} only depend on the logarithmic scale factor $\alpha$, but are all independent of $\pi_{\alpha}$, the change $\Delta\breve\alpha$ in Eq.  \eqref{modifiedscale} vanishes in our case. However, this does not happen for $\Delta\breve\pi_{\alpha}$. From Eq. \eqref{modifiedpi}, at the adopted perturbative order, we get the expression:
\begin{equation}
\label{modifiedpiDEH}
\Delta\breve\pi_{\alpha}= - i\frac{\tilde{M}\omega_k }{2 {\breve \xi}_k^2} e^{{\breve \alpha}} \sum_{\vec{k}\neq \vec{0},(x,y) } \bigg( a_{\vec{k}}^{(x,y)} b_{\vec{k}}^{(x,y)} + {\bar a}_{\vec{k}}^{(x,y)} {\bar b}_{\vec{k}}^{(x,y)}\bigg). 
\end{equation}
Here, ${\breve \xi}_k$ is the result of replacing $\alpha$ directly with $\breve{\alpha}$ in the definition \eqref{xik} of $\xi_k$. Employing this formula and the expression of $H_D$, a direct calculation shows that
\begin{eqnarray}
\label{breveHDDEH}
{\breve H}_D &=& \delta^{\vec{\tau}}_{\vec{0}} H_{\vec{0}} + \sum_{\vec{k}\neq \vec{\tau}} {\breve H}_{\vec{k}},\\
\label{breveHkDEH}
{\breve H}_{\vec{k}} &=& \frac{e^{-{\breve \alpha}}} {2} \sum_{(x,y)} \bigg[ \breve\xi_k \bigg( {\bar a}_{\vec{k}}^{(x,y)}a_{\vec{k}}^{(x,y)}-  a_{\vec{k}}^{(x,y)}{\bar a}_{\vec{k}}^{(x,y)} +{\bar b}_{\vec{k}}^{(x,y)}b_{\vec{k}}^{(x,y)}  - b_{\vec{k}}^{(x,y)}{\bar b}_{\vec{k}}^{(x,y)} \bigg) \nonumber\\
&-& i\frac{\tilde{M}\omega_k }{{\breve \xi}_k ^2} e^{-{\breve \alpha}}{\breve \pi}_{\alpha}  \bigg( a_{\vec{k}}^{(x,y)} b_{\vec{k}}^{(x,y)} + {\bar a}_{\vec{k}}^{(x,y)} {\bar b}_{\vec{k}}^{(x,y)}\bigg)\bigg].
\end{eqnarray} 
This is the Hamiltonian that generates the dynamical evolution of the creation and annihilation variables 
$\{a_{\vec{k}}^{(x,y)},b_{\vec{k}}^{(x,y)},{\bar a}_{\vec{k}}^{(x,y)},{\bar b}_{\vec{k}}^{(x,y)}\}$, as well as that of the fermionic zero-modes if they exist. According to the results of Ref. \cite{uf-flat} commented above, such dynamics would be unitarily implementable in the corresponding Fock representation of the fermionic system if we were to treat the homogeneous geometry as a classical (nonstationary) background.

\section{Hybrid quantization}

We can now quantize our perturbed cosmological model and impose on it the constraints \`a la Dirac, i.e., as operators that annihilate the physical states. We will carry out a hybrid quantization, in which we will adopt specific quantum representations of the homogeneous sector and of the perturbations, the former one based on LQC, although the treatment is easily generalizable to other approaches to the quantization of the homogeneous geometry. On the tensor product of the corresponding representation spaces, we will impose the quantum version of the constraints. These constraints couple the homogeneous and inhomogeneous quantum subsystems of the model, making the quantization nontrivial. More precisely, the coupling occurs in the zero-mode of the Hamiltonian constraint. In addition, we recall that there are four linear perturbative constraints for each nonzero-mode, but these only affect the scalar perturbations, reducing their number of physical degrees of freedom. 

We call $\mathcal{H}_\text{kin}^\text{matt}$ the kinematical Hilbert space on which we represent the zero-mode of the scalar field. For instance, we can choose the Hilbert space $L^2(\mathbb{R},d\phi)$ of square integrable functions over the real line, with $\hat\phi$ acting on it by multiplication, and its canonical momentum $\hat{\pi}_\phi$ acting as $-i\partial_\phi$. On the other hand, we call $\mathcal{H}_\text{kin}^\text{grav}$ the representation space for the homogeneous geometry, that for LQC can be chosen as the space of square summable functions over the points of the real line with the discrete topology \cite{lqc3,lqc4}. Functions of the homogeneous scale factor act by multiplication in this representation. On $\mathcal{H}_\text{kin}^\text{grav}$, we need to represent the variables $(\breve{\alpha},{\breve \pi}_{\alpha})$. Preparing the road to an LQC quantization, we can consider instead the canonical pair $(v,b)$ defined via Eqs. \eqref{alfav} and \eqref{momalfa}, once $(\alpha,\pi_{\alpha})$ has been replaced with $(\breve{\alpha},{\breve \pi}_{\alpha})$ on the left-hand side of those equations. Let $\hat{v}$ and $\hat{b}$ be the associated elementary operators. Denoting $\hat{V}=2\pi\gamma\Delta_g^{1/2}|\hat{v}|$, which is a positive operator, we can then represent $e^{\breve{\alpha}}$ by $[3/(4\pi \sigma)]^{1/3}\hat{V}^{1/3}$. Similarly, we can construct a (self-adjoint, and hence symmetric) operator $\hat{\Omega}_0$ to represent $2\pi\gamma vb$ (the proportionality constant in this expression is standard in LQC). Then, $-3\hat{\Omega}_0/(4\pi\gamma)$ provides an operator version of ${\breve \pi}_{\alpha}$. 

More explicitly, in LQC the kinematical Hilbert space
$\mathcal{H}_\text{kin}^\text{grav}$ can be identified with the span of the basis of eigenstates $|v\rangle$ of $\hat{v}$, with eigenvalue $v\in\mathbb{R}$, taking as inner product the discrete one, $\langle v'|v\rangle=\delta_{v}^{v'}$ \cite{lqc3,lqc4}. The operator $\hat{v}$, determined by the action $\hat{v} |v\rangle=v  |v\rangle$, has then a discrete spectrum. The basic holonomy operators $e^{\pm i \hat{b}/2}$ shift the label of these states in a unit, $e^{\pm i \hat{b}/2} |v\rangle=|v\pm1\rangle$ \cite{lqc1}. Calling $\widehat{\sin(b)}=i( e^{- i \hat{b}}- e^{ i \hat{b}})/2$, and adopting the symmetric ordering proposed in Ref. \cite{MMO}, we can define 
\begin{equation}\label{Omega0}
\hat{\Omega}_0=\frac1{2\sqrt{\Delta_g}}{\hat V}^{1/2}\left[\widehat{{\mathrm{sign}}(v)}\widehat{\sin(b)}+\widehat{\sin(b)}\widehat{{\mathrm{sign}}(v)}\right]{\hat V}^{1/2}.
\end{equation} 

Combining these definitions and choices, we can straightforwardly obtain the quantum representation of the homogeneous contribution $H_{|0}$ to the zero-mode of the Hamiltonian constraint, given in Eq. \eqref{H02}. Leaving aside a global factor of $3e^{-3{\breve \alpha}}/(4\pi)$, that we can absorb with a convenient redefinition of the homogeneous lapse function ${\breve N}_0$, we get the quantum operator $({\hat{\pi}}_{\phi}^2 - \hat{\mathcal H}_0^{(2)})/2$, where
\begin{equation}
\label{H20quantum}
\hat{\mathcal H}_0^{(2)}=\frac{3}{4\pi \gamma^2}\hat\Omega_0^2- \hat V^2 m^2\hat\phi^2.
\end{equation}
The operator  $\hat\Omega_0^2$ annihilates the zero-volume state and leaves invariant its orthogonal complement, without relating the subspaces $\mathcal H_\mathrm{\varepsilon}^\pm$ spanned by states supported on the semilattices $\mathcal L_\mathrm{\varepsilon}^\pm=\{\pm(\varepsilon+4n)|n\in\mathbb N\}$, where $\varepsilon\in(0,4]$ \cite{MMO}. In each of these superselection sectors, the homogeneous volume $v$ has a strictly positive minimum $\varepsilon$ (or a negative maximum $-\varepsilon$). Using these results, we can restrict the discussion of physical states in LQC, e.g., to $\mathcal H_\mathrm{\varepsilon}^+$, corresponding to states with positive $v\in \mathcal L_\mathrm{\varepsilon}^+$. 

Another operator that we will need in our quantization is the regulated version of the inverse of the volume, $\widehat{[1/V]}$. Using standard conventions in LQC, we define it as the cube of the regularized operator
\begin{align}
\widehat{\left[\frac1{V}\right]}^{1/3}=\frac3{4\pi \gamma \sqrt{\Delta_g}}\widehat{{\mathrm{sign}}(v)} \hat V^{1/3}\left[e^{ -i \hat{b}/2} \hat V^{1/3}e^{ i \hat{b}/2}-e^{ i \hat{b}/2}V^{1/3}e^{ -i \hat{b}/2}\right].
\end{align}
This operator is well-defined on the subspaces $\mathcal H_\mathrm{\varepsilon}^\pm$ and commutes with $\hat V$.

Let us now consider the representation of the perturbations. For the linear perturbative constraints and their canonical momenta, we assume a representation in which the mentioned constraints act as (generalized) derivatives. Their quantum imposition then simply implies that the physical states do not depend on this sector of degrees of freedom of the perturbations \cite{hybr-ref}. We can then focus our attention on the rest of perturbative variables: the Mukhanov-Sasaki gauge invariant and its momentum, the tensor perturbations, and the fermion modes. On the system that they form with the homogeneous sector, the only quantum constraint that remains is that corresponding to the zero-mode of the Hamiltonian constraint.

For this part of the perturbative sector, we adopt a tensor product of Fock representations, similar to those discussed for the gauge-invariant scalar, the tensor perturbations, and the Dirac field in Refs. \cite{hybr-inf2,hybr-inf3}, \cite{hybr-ten}, and \cite{uf-flat}, respectively. The Fock spaces for the scalar and tensor perturbations are symmetric, while the fermionic one is antisymmetric. All these Fock representations --or, strictly speaking, a family of unitarily equivalent representations in each case-- have been selected based on our criterion of unitary dynamics of the creation and annihilation variables, and symmetry invariance (as well as a reasonable concept of particles and antiparticles in the fermionic case). We call ${\mathcal F}_s$, ${\mathcal F}_T$, and ${\mathcal F}_D$ the corresponding Fock spaces, where the subindices $s$, $T$, and $D$ refer to scalar perturbations, tensor perturbations, and Dirac fermions, respectively. To simplify the notation, we include in ${\mathcal F}_D$ the Dirac zero-modes, even if we may adopt for them a representation in terms of variables other than creation and annihilation ones. With this convention, a basis of states in each of these Fock spaces is provided by the occupancy-number states $|{\mathcal N} \rangle_s$, $|{\mathcal N} \rangle_T$, and $|{\mathcal N} \rangle_D$, where ${\mathcal N}$ denotes an array of (positive integer) occupancy numbers in each of the considered cases. Creation and annihilation operators (for which we adopt standard conventions and notation) act increasing and decreasing these occupancy numbers, as usual. 

Together with our discussion of the homogeneous sector, we thus conclude that the physical states of our system can be determined starting with elements of the space 
\begin{equation}
\label{represpace}
\mathcal{H}=\mathcal{H}_\text{kin}^\text{grav}\otimes \mathcal{H}_\text{kin}^\text{matt}\otimes {\mathcal F}_s\otimes {\mathcal F}_T\otimes {\mathcal F}_D,
\end{equation}
by imposing the quantum version of the zero-mode of the Hamiltonian constraint. To complete the quantum representation of this constraint, we still have to consider the quadratic contributions of the perturbations. With the redefinition of the lapse function commented above, these contributions are given classically by $4\pi e^{3{\breve \alpha}}(\breve{H}_{|2}+ {^T \tilde{H}_{|2}} + {\breve H}_D)/3$. The only step to reach the desired representation that is not straightforward is the construction of operator versions of the factors that appear multiplying the quadratic powers of the perturbations in this contribution, which are nonlinear functions of the homogeneous variables. For the scalar and tensor parts, we adopt the same prescriptions as in Refs. \cite{hybr-ref,hybr-ten}. Therefore:
i) we take a symmetric multiplicative factor ordering for products of the form $f(\phi)\pi_{\phi}$, ii) we adopt an algebraic symmetrization in factors of the form $V^rg(2\pi\gamma vb)$ for any function $g$ and real number $r$, so that we assign to them the operators $\hat V^{r/2} g(\hat\Omega_0)\hat V^{r/2}$, iii) the same type of algebraic symmetric factor ordering is taken for powers of the inverse volume,  iv) even powers of  $2\pi\gamma vb\propto \pi_{\breve \alpha}$ are represented by the same powers of $\hat \Omega_0$, and v) odd powers $(2\pi\gamma vb)^{2k+1}$, with $k$ an integer, are represented by $|\hat\Omega_0|^k \hat\Lambda_0|\hat\Omega_0|^k$, where $|\hat\Omega_0|=(\hat\Omega_0^2)^{1/2}$ and $\hat{\Lambda}_0$ is defined as $\hat \Omega_0$ but doubling the length of the holonomies, so that the sine operator is replaced with half the sine of the double angle in Eq. \eqref{Omega0}. This doubling of the holonomies length is necessary to leave the superselection sectors  $\mathcal H_\mathrm{\varepsilon}^\pm$ invariant under the action of our constraint. To these prescriptions, we have to add the following for the fermionic contributions: vi) for ${\breve \xi}_{k}$ and any of its algebraic powers (including negative ones), we define the operator representation in terms of $\hat{V}$ using the spectral theorem, so that it commutes with $\hat{V}$, and besides admits (at least locally) a series expansion in powers of $\omega_k$, and vii) in contributions that create or annihilate fermions [arising from the second term in Eq. \eqref{breveHkDEH}], we adopt again an algebraic symmetric ordering for operators of the volume similar to that specified in (ii) and (iii), given by ${\breve \xi}_{k}^{-1}e^{\breve \alpha/2}{\breve \pi}_{\alpha}e^{\breve \alpha/2}{\breve \xi}_{k}^{-1}$, and then adopt the prescriptions explained above. With this procedure, we obtain the representation of the remaining Hamiltonian constraint in our hybrid approach.

The resulting constraint can be expressed as
\begin{equation}
\label{quantumconstraint}
\hat{H}= \frac{1}{2}\big[\hat{\pi}_{\phi}^2-\hat{\mathcal H}_0^{(2)}-\hat{\Theta}_e-\big(\hat\Theta_o\hat\pi_{\phi}\big)_{\rm{sym}}-\hat{\Theta}_T - \delta_{\vec{0}}^{\vec{\tau}} {\hat \Upsilon}_{\vec{0}} - {\hat \Upsilon}_{F}- {\hat \Upsilon}_I \big],
\end{equation}
where we have adopted the symmetrization $(\hat\Theta_o\hat\pi_{\phi})_{\rm{sym}} =(\hat\Theta_o\hat\pi_{\phi}+\hat\pi_{\phi}\hat\Theta_o)/2$ and the different perturbative terms are defined as follows. 
For the scalar gauge invariants, with the notation
\begin{eqnarray}
\label{thetascalar}
\Theta_e=-\sum_{\vec{n},\epsilon} \big[(\vartheta_e \omega_n^2+\vartheta_e^q)v_{\vec n,\epsilon}^2+ \vartheta_e \pi_{v_{\vec n,\epsilon}}^2\big],\qquad
\Theta_o=   - \sum_{\vec{n},\epsilon} \vartheta_o v_{\vec n,\epsilon}^2,
\end{eqnarray}
and our prescriptions for the quantization, we arrive at the following $\vartheta$-operators of only the homogeneous geometry:
\begin{eqnarray}
\label{varthetae}
\hat\vartheta_e&=& l_0 \hat V^{2/3},\\
\label{varthetaeq}
\hat\vartheta_e^q&=&\frac{4\pi }{3 l_0}\widehat{\left[\frac1{V}\right]}^{1/3}\hat{\mathcal H}_0^{(2)}\left(19-24 \pi \gamma^2 \hat\Omega_0^{-2}\hat{\mathcal H}_0^{(2)}\right) \widehat{\left[\frac1{V}\right]}^{1/3}+\frac{m^2}{l_0 }\hat V^{4/3}\left( 1- \frac{8\pi }{3} \hat\phi^2\right),\\
\label{varthetao}
\hat\vartheta_o&=&\frac{16\pi}{l_0}\gamma m^2\hat\phi \hat V^{2/3} |\hat\Omega_0|^{-1} \hat\Lambda_0|\hat\Omega_0|^{-1}\hat V^{2/3}.
\end{eqnarray}
We recall that $l_0=[4\pi/(3\sigma^2)]^{1/3}$. Similarly, for the tensor perturbations, we have \cite{hybr-ten}
\begin{equation}
\Theta_T = -\sum_{\vec {n},\epsilon,\tilde{\epsilon}}\big[\big(\vartheta_e \omega_n^2+ \vartheta_T^q \big)\tilde d_{\vec {n},\epsilon,\tilde{\epsilon}}^2 + \vartheta_e \pi_{\tilde d_{\vec {n},\epsilon,\tilde{\epsilon}}}^2\big],
\end{equation}
and our prescriptions lead to the new $\vartheta$-operator 
\begin{equation}
\label{varthetaqT}
\hat{\vartheta}_T^q=\frac{4\pi}{3l_0}\widehat{\left[\frac1{V}\right]}^{1/3}\hat{\mathcal H}_0^{(2)}\widehat{\left[\frac1{V}\right]}^{1/3}-\frac{8\pi }{3l_0}m^2\hat V^{4/3} \hat{\phi}^2.
\end{equation}
Finally, for the Dirac contribution, including the zero-modes [represented here as operators  $(\hat{m}_{\vec{0}},\hat{r}_{\vec{0}}, \hat{ s}_{\vec{0}}, \hat{ t}_{\vec{0}})$, for instance choosing for them a holomorphic representation similar to that of Ref. \cite{DEH}] and taking normal ordering for all other modes, we adopt the definitions 
\begin{eqnarray}
\label{Upsilonvec0}
\hat{\Upsilon}_{\vec{0}}&=& -2 M \hat{V} \big( \hat{s}_{\vec{0}} \hat{ r}^{\dagger}_{\vec{0}} + \hat{r}_{\vec{0}} \hat{ s}^{\dagger}_{\vec{0}} + \hat{m}_{\vec{0}} \hat{ t}^{\dagger}_{\vec{0}} + \hat{t}_{\vec{0}} \hat{ m}^{\dagger}_{\vec{0}} \big),
\\
\label{UpsilonF}
\hat{\Upsilon}_F &=& - \sum_{\vec{k}\neq\vec{\tau}, (x,y)} 2 l_0  {\breve \xi}_k(\hat{V}) \hat{V}^{2/3} \bigg( \hat{a}_{\vec{k}}^{(x,y)\dagger}\hat{a}_{\vec{k}}^{(x,y) }+\hat{ b}_{\vec{k}}^{(x,y)\dagger}\hat{b}_{\vec{k}}^{(x,y)}  \bigg), \\
\hat{\Upsilon}_I &=& - i \sum_{\vec{k}\neq\vec{\tau}, (x,y)} \frac{ M\omega_k}{l_0\gamma}  {\breve \xi}_k^{-1}(\hat{V}) \hat{V}^{1/6} {\hat \Lambda}_{0} \hat{V}^{1/6} {\breve \xi}_k^{-1}(\hat{V})\nonumber\\
&&\times  \bigg( {\hat a}_{\vec{k}}^{(x,y)} {\hat b}_{\vec{k}}^{(x,y)} + {\hat a}_{\vec{k}}^{(x,y)\dagger} {\hat b}_{\vec{k}}^{(x,y)\dagger}\bigg). 
\label{UpsilonI}
\end{eqnarray}

The generalization of our discussion to a generic potential for the scalar field is rather straightforward. Actually, one only needs to replace any quadratic term of the homogeneous scalar field of the form $m^2\phi^2/2$ with the considered potential, any linear term $m^2\phi$ with the derivative of the potential, and the constant $m^2$ with the second derivative of the potential in every contribution in which it is not accompanied by the homogeneous scalar field. Finally, we can also generalize the analysis to hybrid quantizations in which the homogeneous geometry is not quantized \`a la loop. To use our formulas in any other representation of the geometry, we only have to identify the operators that play in it the role of $\hat{V}$, its regularized inverse (which might coincide with the true inverse), $\hat{\Omega}_0$, and its modified version $\hat{\Lambda}_0$. In this way, our results can be applied to a variety of schemes other than LQC.

\section{Born-Oppenheimer approximation}

We will now introduce an ansatz describing a family of physical states that are interesting in realistic scenarios, and in particular in situations in which the perturbations are not expected to affect much the homogeneous geometry. We seek physical states in which the dependence on this homogeneous geometry, on the scalar perturbations, on the tensor ones, and on fermions, can be separated. The homogeneous scalar field $\phi$ will be regarded as an internal time, so that each part of the wave function $\Xi$ of our physical states may depend on it. With our notation for the occupancy numbers of scalar (s), tensor (T), and Dirac (D) perturbations, and denoting the dependence on the homogeneous geometry symbolically with a dependence on $V$, our ansatz can be expressed 
\begin{equation}\label{Xi}
\Xi=\Gamma(V,\phi) \psi({\mathcal N_s},{\mathcal N_T},{\mathcal N_D},\phi)= \Gamma(V,\phi) \psi_s({\mathcal N_s},\phi)\psi_T({\mathcal N_T},\phi)\psi_D({\mathcal N_D},\phi).
\end{equation}
The homogeneous part is chosen as a wave function of the form \cite{hybr-ref,hybr-inf5}
\begin{equation}\label{Gammahomo}
\Gamma(V,\phi)= \hat{U}_0(V,\phi)\chi(V),
\end{equation}
where $\hat{U}_0$ is a unitary evolution operator in the internal time $\phi$, that we suppose generated by a self-adjoint operator, defined as $\hat{\tilde{\mathcal H}}_0=[\hat\pi_{\phi},\hat{U}_0]  {\hat{U}_0^{-1} }$. The state $\Gamma$ can be considered as a solution of the homogeneous part of the constraint up to the order of the perturbative contributions. For this, we assume that the difference $ (\hat{\tilde{\mathcal H}}_0)^2  -\hat{\mathcal H}_0^{(2)}$ is negligibly small on $\Gamma$ at all orders dominant over the quadratic one in the perturbations. In addition to this, either we have that $[\hat{\pi}_{\phi}, \hat{\tilde{\mathcal H}}_0]$ is also negligible up to second perturbative order, included, or it is most convenient to absorb this commutator by slightly changing the factor ordering in the homogeneous part of the constraint \eqref{quantumconstraint} \cite{hybr-inf5}. For instance, we can adopt the ordering\footnote{Factor orderings of this kind can be related with the definition of the state $\Gamma$ in the unperturbed system by means of group averaging techniques \cite{marolf,marolf2}.} $(\hat{\pi}_{\phi}+\hat{\tilde{\mathcal H}}_0) (\hat{\pi}_{\phi}-\hat{\tilde{\mathcal H}}_0)  +  \{(\hat{\tilde{\mathcal H}}_0)^2  -\hat{\mathcal H}_0^{(2)}\}$, so that its action on the state \eqref{Gammahomo} coincides with the action of the last contribution between curved brackets, which is at most of quadratic perturbative order according to our assumptions. Finally, we also assume that $\hat{\tilde{\mathcal H}}_0$ is positive. In fact, a suggested operator is the square root of the positive part of $\hat{\mathcal H}_0^{(2)}$ \cite{hybr-inf5}, although we will not restrict our discussion here to a specific choice. As for the state $\chi$ above, we take it normalized with respect to the inner product of the homogeneous geometry, i.e. in ${\mathcal H}_\mathrm{kin}^\mathrm{grav}$. We can think of $\chi$ as the initial state for the homogeneous geometry, and it 
 would be natural to choose it with a highly semiclassical behavior, strongly peaked on a certain FLRW geometry.

With our ansatz and mild hypotheses, the constraint equation on $\Xi$ becomes 
\begin{eqnarray}
\label{constraintBO}
\Gamma (\hat{\pi}_{\phi}^2\psi)&+&2 (\hat{\tilde{\mathcal H}}_0\Gamma) (\hat{\pi}_{\phi}\psi)+\Big(\Big\{ (\hat{\tilde{\mathcal H}}_0)^2- \hat{\mathcal H}_0^{(2)}\Big\} \Gamma\Big)\psi+\frac{i}{2}{\mathrm d}_{\phi}\hat{\Theta}_o(\Gamma\psi)-\hat{\Theta}_o\big\{\Gamma(\hat{\pi}_{\phi}\psi)  \big\}\nonumber\\
&-&\big\{\hat{\Theta}_e+(\hat{\Theta}_o\hat{\tilde{\mathcal H}}_0)_{\mathrm{sym}}+\hat{\Theta}_T + \delta_{\vec{0}}^{\vec{\tau}} {\hat \Upsilon}_{\vec{0}} + {\hat \Upsilon}_{F}+ {\hat \Upsilon}_I \big\}(\Gamma\psi)=0,
\end{eqnarray}
where we have used the notation $-i{\mathrm d}_{\phi}\hat O\equiv [\hat{\pi}_{\phi}-\hat{\tilde{\mathcal H}}_0,\hat O]$, with $\hat O$ being a generic operator. In the above constraint equation, all the dependence on $\hat{\pi}_{\phi}$ has been shown explicitly. In this sense, note that, with our definitions, ${\mathrm d}_{\phi}\hat{\Theta}_o$ is independent of this momentum. 

Let us now introduce the assumption that, on states of the Born-Oppenheimer type, one can ignore any quantum transition in the homogeneous geometry mediated by the constraint \cite{hybr-ref}. If this is the case, the constraint equation is tantamount to taking its expectation value on the homogeneous state $\Gamma$. One can prove that this assumption holds if and only if one can neglect the dispersions on $\Gamma$, relative to the corresponding expectation values, of the operators $\hat{\tilde{\mathcal H}}_0$ and $(\hat{\tilde{\mathcal H}}_0)^2- \hat{\mathcal H}_0^{(2)}$, as well as of $\hat{\vartheta}_e$ and $\hat{\vartheta}_e^q+(\hat{\vartheta}_o\hat{\tilde{\mathcal H}}_0)_{\mathrm{sym}}-i{\mathrm d}_{\phi}\hat{\vartheta}_o/2$ in the presence of scalar perturbations, of $\hat{\vartheta}_T^q$ if there are also tensor perturbations, of $\hat{V}$ if there are Dirac zero-modes, and, finally, of $\hat{V}^{2/3}{\breve \xi}_k(\hat{V})$ and ${\breve \xi}_k^{-1}(\hat{V}) \hat{V}^{1/6} {\hat \Lambda}_{0} \hat{V}^{1/6} {\breve \xi}_k^{-1}(\hat{V})$ in the presence of other modes of the Dirac field. Remarkably, in the absence of fermionic nonzero-modes, the number of operators of the homogeneous geometry that must be peaked on $\Gamma$ is finite (and small in number, in fact), in spite of the existence of an infinite number of degrees of freedom in the system. It is only the introduction of fermions that puts the classicality of the quantum state of the homogeneous geometry to a severe test, since their presence, and the nonconformal coupling with the geometry that their mass involves, requires a peaked behavior of an infinite number of operators. At least, we point out that the dependence of all these operators on $\hat{\Lambda}_0$ is the same, and the only change is in their dependence on the volume. 

If our assumption is valid, and we denote the expectation value of a generic operator $\hat{O}$ on $\Gamma$ by $\langle \hat{O} \rangle_{\Gamma}$, we arrive at
\begin{eqnarray}
\label{constraintBO2}
\hat{\pi}_{\phi}^2\psi &+& 2 \langle \hat{\tilde{\mathcal H}}_0 \rangle_{\Gamma} \hat{\pi}_{\phi} \psi + \langle (\hat{\tilde{\mathcal H}}_0)^2- \hat{\mathcal H}_0^{(2)} \rangle_{\Gamma} \psi -\langle \hat{\Theta}_e+(\hat{\Theta}_o\hat{\tilde{\mathcal H}}_0)_{\mathrm{sym}} \rangle_{\Gamma} \psi \nonumber \\
\, &+& \frac{i}{2} \langle {\mathrm d}_{\phi} \hat{\Theta}_o \rangle_{\Gamma} \psi -\langle \hat{\Theta}_T \rangle_{\Gamma} \psi - \langle \delta_{\vec{0}}^{\vec{\tau}} \hat{\Upsilon}_{\vec{0}} + {\hat \Upsilon}_{F} + {\hat \Upsilon}_I \rangle_{\Gamma} \psi = 0.
\end{eqnarray}
In agreement with our perturbative approximations, we have neglected a term $\langle\hat{\Theta}_o\rangle_{\Gamma}\hat{\pi}_{\phi}\psi$ compared to the second contribution in the above equation. Besides, we note that our expectation values are taken only over the homogeneous geometry, i.e. with respect to the inner product in $\mathcal H_\mathrm{kin}^\mathrm{grav}$, and that the result, in general, is an operator defined  on $\mathcal H_\mathrm{kin}^\mathrm{matt}\otimes \mathcal{F}_s\otimes \mathcal{F}_T\otimes \mathcal{F}_D$.

With our Born-Oppenheimer ansatz, that separates the dependence on the scalar gauge invariants, the tensor perturbations, and the Dirac fermions, the constraint equation \eqref{constraintBO2} leads to Schr\"odinger equations for each of these perturbative sectors under certain reasonable hypotheses \cite{hybr-ref}. The most important of these hypotheses is that $\hat{\pi}_{\phi}^2 \psi$ must be negligible compared to the rest of terms, and in particular in comparison with the term that is proportional to $\hat{\pi}_{\phi} \psi$. Essentially, this condition requires that the contribution of the wave funtion of the perturbations to the momentum of the homogeneous scalar field be much smaller than its value on $\Gamma$. The consistency of this hypothesis can be confirmed, once $\hat{\pi}_{\phi} \psi$ is estimated using the assumption and the constraint, by taking the derivative of the result with respect to $\phi$ and comparing the neglected quantity with $\langle \hat{\tilde{\mathcal H}}_0 \rangle_{\Gamma} \hat{\pi}_{\phi} \psi $. The other, much less relevant hypothesis leading to Scr\"odinger equations is that one can neglect the contribution of $\langle {\mathrm d}_{\phi} \hat{\Theta}_o \rangle_{\Gamma}$. Note that this contribution affects only the scalar part of the perturbations, and therefore only their Schr\"odinger equation. The hypothesis is necessary inasmuch as one requires a unitary evolution for the Mukhanov-Sasaki modes, otherwise one can proceed keeping the corresponding term in our considerations. 

Accepting these two hypotheses, and recalling that we have assumed that $\hat{\tilde{\mathcal H}}_0$ is defined as a positive operator, so that in particular it is meaningful to divide by its expectation value, one arrives at the following evolution equations in $\phi$ for the different perturbative sectors:
\begin{eqnarray}
\label{schroscalar}
\hat{\pi}_{\phi}\psi_s&=&\left[\frac{\langle \hat{\Theta}_{e}+ \big(\hat{\Theta}_{o} \hat{\tilde{\mathcal H}}_0\big)_{\mathrm{sym}}\rangle_{\Gamma}}{2 \langle \hat{\tilde{\mathcal H}}_0 \rangle_{\Gamma}}+C_s^{(\Gamma)}(\phi)\right]\psi_s,
\\
\label{schrotensor}
\hat{\pi}_{\phi}\psi_T&=&\left[\frac{\langle \hat{\Theta}_{T} \rangle_{\Gamma}}{2 \langle \hat{\tilde{\mathcal H}}_0 \rangle_{\Gamma}}+C_T^{(\Gamma)}(\phi)\right]\psi_T,
\\
\label{schrofermion}
\hat{\pi}_{\phi}\psi_D&=&\left[\frac{\langle \delta_{\vec{0}}^{\vec{\tau}} \hat{\Upsilon}_{\vec{0}} + {\hat \Upsilon}_{F} + {\hat \Upsilon}_I  \rangle_{\Gamma}}{2 \langle \hat{\tilde{\mathcal H}}_0 \rangle_{\Gamma}}+C_D^{(\Gamma)}(\phi)\right]\psi_D,
\end{eqnarray}
where $C_s^{(\Gamma)}(\phi)$, $C_T^{(\Gamma)}(\phi)$, and $C_D^{(\Gamma)}(\phi)$ are functions of $\phi$, possibly dependent on the state $\Gamma$ considered for the homogeneous geometry, such that
\begin{equation}
\label{Gammaconstants}
C_s^{(\Gamma)}(\phi)+C_T^{(\Gamma)}(\phi)+C_D^{(\Gamma)}(\phi)=\langle (\hat{\tilde{\mathcal H}}_0)^2 - \hat{\mathcal H}_0^{(2)} \rangle_\Gamma.
\end{equation}
In some sense, this quantity can be interpreted as a backreaction on the homogeneous geometry, inasmuch as $\langle (\hat{\tilde{\mathcal H}}_0)^2 - \hat{\mathcal H}_0^{(2)} \rangle_\Gamma$ indicates a departure of the homogeneous state $\Gamma$ from an exact solution to the zero-mode of the Hamiltonian constraint in the absence of perturbations, case in which the left-hand side of Eq. \eqref{Gammaconstants} vanishes. Note, nonetheless, that in principle one can have no departure at all if the backreaction of the fermions is counterbalanced by the contribution of the scalar and tensor perturbations.

Before closing this section, let us comment that one can derive effective equations for the perturbations from our previous discussion paralleling the arguments explained in Ref. \cite{hybr-ref}. One can extract them from Eq. \eqref{constraintBO2}, assuming that the quadratic dependence on $\hat{\pi}_{\phi}$ and on the operators that describe the degrees of freedom of the perturbations has an associated effective dynamics on the considered physical state, dynamics that is obtained essentially by replacing those operators by their direct classical counterpart. Alternatively, one can admit the validity of the hypotheses necessary for the above Scr\"odinger equations and consider the dynamics generated by the respective Hamiltonians in those equations, accepting that the perturbative operators that appear in them can be treated effectively as classical variables. We refer the reader to Refs. \cite{hybr-ref,hybr-inf5} for further details on this topic.

\section{Quantum dynamics of the fermionic perturbations}

Let us now discuss the quantum dynamics of the creation and annihilation operators for the nonzero-modes of the Dirac field that follows from the Schr\"odinger Eq.\eqref{schrofermion} or, alternatively, directly from the quantum constraint \eqref{constraintBO2}, if one neglects the contribution of the perturbations to the momentum of the homogeneous scalar field in comparison with the homogeneous contribution $\langle \hat{\tilde{\mathcal H}}_0 \rangle_{\Gamma}$. It is straightforward  to see that the resulting evolution equations are
\begin{eqnarray}
\label{equationmotiona}
d_{\eta_\Gamma}{\hat a}_{\vec{k}}^{(x,y)}(\eta,\eta_0)&=&- i F_k^{(\Gamma)}{\hat a}_{\vec{k}}^{(x,y)}(\eta,\eta_0)+ G_k^{(\Gamma)} {\hat b}_{\vec{k}}^{(x,y)\dagger}(\eta,\eta_0),\nonumber
\\
d_{\eta_\Gamma}{\hat b}_{\vec{k}}^{(x,y)\dagger}(\eta,\eta_0)&=& i F_k^{(\Gamma)}{\hat b}_{\vec{k}}^{(x,y)\dagger}(\eta,\eta_0)- G_k^{(\Gamma)} {\hat a}_{\vec{k}}^{(x,y)}(\eta,\eta_0),
\end{eqnarray}
where, by convenience, we have introduced a conformal time $\eta_{\Gamma}$, that is defined in terms of the homogeneous scalar field by means of the relation\footnote{Note that classically this time would coincide with the conformal time introduced below Eq. \eqref{1order}.}
\begin{equation}
\label{etaGamma}
d{\eta_\Gamma}=  \frac{l_0\langle \hat{V}^{2/3} \rangle_{\Gamma} }{ \langle \hat{\tilde{\mathcal H}}_0 \rangle_{\Gamma} } d\phi,
\end{equation}
We are evaluating it at the instant $\eta$, and $F_k^{(\Gamma)}$ and $G_k^{(\Gamma)}$ are the following mode-dependent functions of time:
\begin{eqnarray}
\label{FkGamma}
F_k^{(\Gamma)}&=&\frac{\langle  {\breve \xi}_k(\hat{V}) \hat{V}^{2/3} \rangle_{\Gamma} }{\langle \hat{V}^{2/3} \rangle_{\Gamma}},
\\
\label{GkGamma}
G_k^{(\Gamma)}&=& \frac{ M \omega_k}{2l_0^2} \,\frac{\langle  {\breve \xi}_k^{-1}(\hat{V}) \hat{V}^{1/6} \hat{\Lambda}_0 \hat{V}^{1/6} {\breve \xi}_k^{-1}(\hat{V}) \rangle_{\Gamma} }{\gamma\langle \hat{V}^{2/3} \rangle_{\Gamma}}.
\end{eqnarray}
Note that the dependence of these functions on the mode is only through $\omega_k$, and not through the rest of details of the specific tuple $\vec{k}$ under consideration. Besides, all our definitions include an implicit dependence on the particular state $\Gamma$ considered for the homogeneous geometry. In addition, in all these expressions, the dependence on the conformal time appears via the dependence on the homogeneous field $\phi$, including the dependence of $\Gamma$, once the relation \eqref{etaGamma} has been integrated. Finally,  for our evolution equations, we take the operators ${\hat a}_{\vec{k}}^{(x,y)}$ and ${\hat b}_{\vec{k}}^{(x,y)\dagger}$ as initial conditions at an arbitrary initial time $\eta_0=\eta_{\Gamma}(\phi_0)$. 

It is worth remarking that, since $\hat{\tilde{\mathcal H}}_0$ is a positive operator by assumption, and since $\hat{V}$ is bounded from below by a positive number in any superselection sector of LQC \cite{MMO}, our change to conformal time and the definition of the functions \eqref{FkGamma} and \eqref{GkGamma} are well-defined. For other possible representations of the homogeneous geometry, like in geometrodynamics, the volume operator might reach a vanishing expectation value, for instance in the big bang for semiclassical states, and might pose intrinsic obstructions to the above constructions.

We expect that the studied dynamics for the nonzero-modes of the Dirac field can be implemented unitarily in our quantum theory, given our choice of Fock representation and the hybrid approach that we have adopted. Nonetheless, the functions $F_k^{(\Gamma)}$ and $G_k^{(\Gamma)}$ that determine the dynamics are not defined by a classical background geometry, but are ratios of expectation values on a quantum state. We could consider situations in which these expectation values are not associated with a semiclassical or effective trajectory. To cope with these issues, we will analyze the quantum dynamics in detail. In the rest of this section, we will study the Bogoliubov transformation that relates the creation and annihilation operators for the nonzero-modes with the operators that represent their initial values. We will leave to the next sections the determination of the operator that implements this  transformation and the proof that it is indeed unitary. With this operator at hand, we will be able to construct solutions to the Schr\"odinger equation of (the nonzero-modes of) the Dirac field. 

Let us start by introducing the following definition of operators, motivated by the classical relation \eqref{avariable}, or rather by its inverse, that can be easily calculated using Eqs. \eqref{g1,g2} and \eqref{f2}, particularized to the case of real coefficients:
\begin{eqnarray}
\label{quantumx}
{\hat x}_{\vec{k}}(\eta,\eta_0)&=& f_{1,k}^{(\Gamma)}{\hat a}_{\vec{k}}^{(x,y)}(\eta,\eta_0)+ f_{2,k}^{(\Gamma)} {\hat b}_{\vec{k}}^{(x,y)\dagger}(\eta,\eta_0),
\\
\label{quantumy}
{\hat y}_{-\vec{k}-2\vec{\tau}}^{\dagger}(\eta,\eta_0)&=& f_{2,k}^{(\Gamma)}{\hat a}_{\vec{k}}^{(x,y)}(\eta,\eta_0)- f_{1,k}^{(\Gamma)} {\hat b}_{\vec{k}}^{(x,y)\dagger}(\eta,\eta_0),
\end{eqnarray}
with $|f_{1,k}^{(\Gamma)}|^2+|f_{2,k}^{(\Gamma)}|^2=1$. Note that we restrict these functions to depend on $\omega_k$, rather than on $\vec{k}$, and to coincide for the possible values of $(x,y)$, namely $(m,s)$ and $(t,r)$. Inspired by the choice made in Eq. \eqref{DEHfs} and the definition of $F_k^{(\Gamma)}$, we take 
\begin{equation}
\label{quantumDEHfs}
f_{1,k}^{(\Gamma)}= \sqrt{ \frac{ F_k^{(\Gamma)} - \omega_k }{ 2 F_k^{(\Gamma)} } } , \qquad f_{2,k}^{(\Gamma)} = \sqrt{  \frac{ F_k^{(\Gamma)} + \omega_k }{ 2 F_k^{(\Gamma)} } }.
\end{equation}
Since $\xi_k(\hat{V}) \geq \omega_k$ as an operator, and $\hat{V}$ is strictly positive, it is ensured that $ F_k^{(\Gamma)}\geq \omega_k$. Hence, $f_{1,k}^{(\Gamma)}$ and $f_{2,k}^{(\Gamma)}$ are well-defined for any state $\Gamma$ and they are real functions. The dynamical equations \eqref{equationmotiona} then translate into
\begin{eqnarray}
\label{equationmotionx}
d_{\eta_\Gamma}{\hat x}_{\vec{k}}(\eta,\eta_0)&=& i \omega_k {\hat x}_{\vec{k}}(\eta,\eta_0)+ H_k^{(\Gamma)} {\hat y}_{-\vec{k}-2\vec{\tau}}^{\dagger}(\eta,\eta_0),
\\
\label{equationmotiony}
d_{\eta_\Gamma}{\hat y}_{-\vec{k}-2\vec{\tau}}^{\dagger}(\eta,\eta_0)&=& -i \omega_k 
{\hat y}_{-\vec{k}-2\vec{\tau}}^{\dagger}(\eta,\eta_0) - \bar{H}_k^{(\Gamma)} {\hat x}_{\vec{k}}(\eta,\eta_0),
\end{eqnarray}
where we have defined
\begin{equation}
\label{HkGamma}
H_k^{(\Gamma)} = - G_k^{(\Gamma)} - i \sqrt{\big(F_k^{(\Gamma)}\big)^2-\omega_k^2} +\frac{ \omega_k \Big( F_k^{(\Gamma)} \Big)^{\prime}} { 2 F_k^{(\Gamma)} \sqrt{\big(F_k^{(\Gamma)}\big)^2-\omega_k^2}},
\end{equation}
and the prime denotes from now on the derivative with respect to the conformal time $\eta_\Gamma$. Recalling that $\hat{\Lambda}_0$ and $\hat{V}$ are representations of $-4\pi\gamma \pi_{\breve{\alpha}}/3$ and $4\pi\sigma e^{3\breve \alpha}/3$, respectively, and that on classical trajectories  $\pi_{\breve{\alpha}}=- {\breve \alpha}^{\prime} e^{2\breve{\alpha}}$ ignoring perturbative corrections, one can check that, on states peaked on such trajectories, the dynamical equations \eqref{equationmotionx} and \eqref{equationmotiony} reproduce the classical ones \eqref{1order}.

Let us now call $\hat{z}_{\vec{k}}$ any of the operators $\hat{x}_{\vec{k}}$ or $\hat y_{\vec{k}}$ indifferently, and define $\hat{\tilde{z}}_{\vec{k}}=(iH_k^{(\Gamma)})^{-1/2} \hat{z}_{\vec{k}}$. It is easy to derive the associated equation
\begin{equation}\label{zequation}
\hat{\tilde{z}}_{\vec{k}}^{\prime\prime}=- \bigg[ \tilde{\omega}_k^2 + \Big|H_k^{(\Gamma)}\Big|^2 + \frac{1}{2} \Big(\ln{H_k^{(\Gamma)}}\Big)^{\prime\prime}\bigg] \hat{\tilde{z}}_{\vec{k}},
\end{equation}
where $\tilde{\omega}_k=\omega_k +i (\ln{H_k^{(\Gamma)}})^{\prime}/2$. As in Appendix B of Ref. \cite{uf1}, we search for two independent solutions of the corresponding classical equation with the form $\tilde{z}_{\vec{k}}^{l}=\exp{[-i(-1)^l\Theta_k^l]}$, where $l=1,2$. We get
\begin{equation}
\label{Thetaasymptotics}
\Theta_k^l=\omega_k(\eta-\eta_0)+\frac{i}{2} \Big[(-1)^l+1\Big]  \ln{\left( \frac{H_k^{(\Gamma)}} {H_k^{(\Gamma),0} } \right) }  +\int_{\eta_0}^{\eta}  \Lambda_k^{l}(\eta_{\Gamma}) d\eta_{\Gamma},
\end{equation}
where we have employed the notation $H_k^{(\Gamma),0}=H_k^{(\Gamma)}(\eta_0)$ and $\Lambda_k^{l}$ is a solution to the Ricatti equation
\begin{eqnarray}
\label{lambdaequation}
\Big( \Lambda_k^l \Big)^{\prime} &=& i (-1)^l  \Big[ \Big(\Lambda_k^l \Big)^2 +2 \tilde{\omega}_k\Lambda_k^l \Big] - u^l_k, \\
\label{uele}
u^l_k &=& i (-1)^l  \Big\vert H_k^{(\Gamma)} \Big\vert^2 +\frac{i}{2} \Big[(-1)^l+1\Big]  \Big( \ln{H_k^{(\Gamma)}} \Big)^{\prime\prime}
\end{eqnarray}
with initial condition $\Lambda_k^l(\eta_0)=0$. An asymptotic analysis like the one carried out in Ref. \cite{uf1} shows then that 
\begin{equation}
\int_{\eta_0}^{\eta}  \Lambda_k^{l}(\eta_{\Gamma}) d\eta_{\Gamma}=
- (-1)^l\frac{i }{2\omega_k}  \int_{\eta_0}^{\eta} u^{l}_k(\eta_{\Gamma})  d\eta_{\Gamma} + \mathcal{O}(\omega_k^{-2}).
\label{Lambdaasymptotics}
\end{equation}
These results are valid under relatively mild conditions on $H_k^{(\Gamma)}$, for instance that it does not vanish and has a fourth-order time derivative that is continuous in the considered interval (so that, in our analysis, all the performed integrations by parts are well-defined). 

The operators $\hat{x}_{\vec{k}}$ or $\hat{y}_{\vec{k}}$ are given by linear combinations of our two independent solutions to Eq. \eqref{zequation}. The coefficients in these combinations are operators that carry the information about the initial conditions. Combining these linear relations with Eqs. \eqref{quantumx} and \eqref{quantumy}, we can express our fermionic creation and annihilation operators as linear combinations of their initial values at $\eta_0$:
\begin{eqnarray}
\label{quantumBogoliubov}
{\hat a}_{\vec{k}}^{(x,y)}(\eta,\eta_0)&=& \alpha_k(\eta,\eta_0){\hat a}_{\vec{k}}^{(x,y)}+\beta_k(\eta,\eta_0){\hat b}_{\vec{k}}^{(x,y)\dagger},\nonumber\\
{\hat b}_{\vec{k}}^{(x,y)\dagger}(\eta,\eta_0)&=&-\bar{\beta}_k(\eta,\eta_0){\hat a}_{\vec{k}}^{(x,y)}+\bar{\alpha}_k(\eta,\eta_0){\hat b}_{\vec{k}}^{(x,y)\dagger}.
\end{eqnarray}
A detailed calculation of the coefficients of this Bogoliubov transformation \cite{uf1} leads finally to the following formula (where we obviate the time dependence and the limits of integration over $(\eta,\eta_0)$ to simplify the notation):
\begin{eqnarray}
\label{alpha}
\alpha_{k}&=&
\bigg[f_{1,k}^{(\Gamma)} \Big(f_{1,k}^{(\Gamma),0}-f_{2,k}^{(\Gamma),0} \zeta_k\Big) e^{i\int \Lambda_k^1} - f_{2,k}^{(\Gamma)}f_{1,k}^{(\Gamma),0} \bar{\zeta}_k \frac{\bar{H}_k^{(\Gamma)}}{\bar{H}_k^{(\Gamma),0}} e^{i\int\bar{\Lambda}_k^2}\bigg] e^{i\omega_k(\eta-\eta_0)}\nonumber\\  &+&
\bigg[f_{2,k}^{(\Gamma)}\Big(f_{1,k}^{(\Gamma),0}\bar{\zeta}_k+f_{2,k}^{(\Gamma),0} \Big) e^{-i\int \bar{\Lambda}_k^1} + f_{1,k}^{(\Gamma)}f_{2,k}^{(\Gamma),0} \zeta_k \frac{H_k^{(\Gamma)}}{H_k^{(\Gamma),0}} e^{-i\int \Lambda_k^2}\bigg] e^{-i\omega_k(\eta-\eta_0)},\\
\beta_{k}&=& \bigg[f_{1,k}^{(\Gamma)}\Big(f_{2,k}^{(\Gamma),0}+f_{1,k}^{(\Gamma),0} \zeta_k\Big) e^{i\int \Lambda_k^1} - f_{2,k}^{(\Gamma)} f_{2,k}^{(\Gamma),0} \bar{\zeta}_k \frac{\bar{H}_k^{(\Gamma)}}{\bar{H}_k^{(\Gamma),0}} e^{i\int\bar{\Lambda}_k^2}\bigg] e^{i\omega_k(\eta-\eta_0)}\nonumber\\  &+&
\bigg[f_{2,k}^{(\Gamma)}\Big(f_{2,k}^{(\Gamma),0}\bar{\zeta}_k-f_{1,k}^{(\Gamma),0} \Big) e^{-i\int \bar{\Lambda}_k^1} - f_{1,k}^{(\Gamma)}f_{1,k}^{(\Gamma),0} \zeta_k \frac{H_k^{(\Gamma)}}{H_k^{(\Gamma),0}} e^{-i\int \Lambda_k^2}\bigg] e^{-i\omega_k(\eta-\eta_0)}.
\label{beta}
\end{eqnarray}
We have used that $f_{1,k}^{(\Gamma)}$ and $f_{2,k}^{(\Gamma)}$ are real, and their initial values at $\eta_0$ have been called $f_{1,k}^{(\Gamma),0}$ and $f_{2,k}^{(\Gamma),0}$, respectively. In addition, we have defined
\begin{equation}
\label{zeta}
\zeta_k=  \frac{iH_k^{(\Gamma),0}}{2\omega_k+i\Big(\ln{H_k^{(\Gamma)}}\Big)^{\prime}_0},
\end{equation}
where the subscript in the derivative of the logarithm stands for evaluation at the initial time. Finally, we notice that, since the canonical anticommutation relations hold at all times, we have that $|\alpha_k(\eta,\eta_0)|^2+|\beta_k(\eta,\eta_0))|^2=1$. 

\section{Evolution operator for the fermion perturbations}

We will discuss now the implementability of the Bogoliubov transformation that encodes the quantum dynamics of the fermionic nonzero-modes in terms of an evolution operator $\hat{U}_D$. 

Based on Eq. \eqref{DEHfs}, we expect $f_{2,k}^{(\Gamma)}$ to be asymptotically of order unity for large $\omega_k$, and $f_{1,k}^{(\Gamma)}$ to be negligible. As a result, the dominant contribution to $\alpha_k(\eta,\eta_0)$ in this asymptotic limit should be given by the term that contains the product $f_{2,k}^{(\Gamma)}f_{2,k}^{(\Gamma),0}$. Recalling in addition Eq. \eqref{Lambdaasymptotics}, we expect that $\alpha_k(\eta,\eta_0)$ behaves asymptotically as the phase $e^{-i\omega_k(\eta-\eta_0)}$. It is most convenient to absorb this phase in a trivial unitary evolution operator $\hat{U}_L$, and deal separately with the remaining Bogoliubov transformation. More specifically, we will adopt the splitting $\hat{U}_D=\hat{U}_B\hat{U}_L$, where $\hat{U}_L$ changes the annihilation operators of the nonzero-modes by a phase $e^{-i\omega_k(\eta-\eta_0)}$ (and the creation operators by the inverse phase), and $\hat{U}_B$ implements the supplementary Bogoliubov transformation with coefficients
\begin{equation}
\label{tildeBogoliubov}
{\tilde \alpha}_k(\eta,\eta_0)=e^{i\omega_k(\eta-\eta_0)}\alpha_k(\eta,\eta_0),\qquad   {\tilde \beta}_k(\eta,\eta_0)=e^{-i\omega_k(\eta-\eta_0)}\beta_k(\eta,\eta_0).
\end{equation}

It is easy to construct the evolution operator $\hat{U}_L$. Defining 
\begin{equation}
\label{Tfree}
\hat{T}_L(\eta,\eta_0)= i(\eta-\eta_0) \sum_{\vec{k}\neq\vec{\tau},(x,y)} \omega_k  \Big( {\hat a}_{\vec{k}}^{(x,y)\dagger} {\hat a}_{\vec{k}}^{(x,y)} +  {\hat b}_{\vec{k}}^{(x,y)\dagger}  {\hat b}_{\vec{k}}^{(x,y)} \Big),
\end{equation}
we simply have $\hat{U}_L= e^{-\hat{T}_L}$. Here and in the following, we avoid displaying the time dependence explicitly, unless necessary, to simplify the notation. Then, we clearly have 
\begin{eqnarray}
\label{quantumBogliubovfree}
\hat{U}_L^{-1} {\hat a}_{\vec{k}}^{(x,y)}\hat{U}_L&=&e^{-i\omega_k(\eta-\eta_0)} {\hat a}_{\vec{k}}^{(x,y)},\nonumber \\
\hat{U}_L^{-1} {\hat b}_{\vec{k}}^{(x,y)\dagger}\hat{U}_L&=&e^{i\omega_k(\eta-\eta_0)}{\hat b}_{\vec{k}}^{(x,y)\dagger}.
\end{eqnarray}

In turn, for the remaining Bogoliubov tansformation, we adopt the parametrization
\begin{eqnarray}
\label{alphaparametrized}
{\tilde \alpha}_k&=&\cos{\sqrt{|\Delta_k|^2+\rho_k^2}} +i \rho_k \frac{\sin{\sqrt{|\Delta_k|^2+\rho_k^2}}}{\sqrt{|\Delta_k|^2+\rho_k^2}},
\nonumber \\
{\tilde \beta}_k&=&-\Delta_k \frac{\sin{\sqrt{|\Delta_k|^2+\rho_k^2}}}{\sqrt{|\Delta_k|^2+\rho_k^2}},
\end{eqnarray}
where $\rho_k$ is a real number, and $\Delta_k$ is complex. Associated to this Bogoliubov transformation, we can introduce the quadratic operator
\begin{eqnarray}
\label{quantumT}
\hat{T}_B&=&\sum_{\vec{k}\neq\vec{\tau},(x,y)} \!\!\Big[ \Delta_k {\hat a}_{\vec{k}}^{(x,y)\dagger} {\hat b}_{\vec{k}}^{(x,y)\dagger}  - \bar{\Delta}_k {\hat b}_{\vec{k}}^{(x,y)} {\hat a}_{\vec{k}}^{(x,y)}\nonumber \\
&-& i \rho_k \Big( {\hat a}_{\vec{k}}^{(x,y)\dagger} {\hat a}_{\vec{k}}^{(x,y)} +  {\hat b}_{\vec{k}}^{(x,y)\dagger}  {\hat b}_{\vec{k}}^{(x,y)} \Big) + i c_k^{(x,y)} \Big],
\end{eqnarray}
where $c_k^{(x,y)}$ is a time-dependent (c-number) phase that we leave arbitrary for the moment. We next define $\hat{U}_B= e^{-\hat{T}_B}$. Then, employing the formula
\begin{equation}
\label{actionU}
e^{\hat{T}_{B}} \hat{O} e^{-\hat{T}_{B}}= \hat{O}+ \sum_{n=1}^{\infty}\frac{1}{n!} [\hat{T}_{B},...[\hat{T}_{B},\hat{O}]]_{(n)},
\end{equation}
where $[. , .]_{(n)}$ denotes the $n$th commutator and $\hat{O}$ is a generic operator, and recalling Eq. \eqref{alphaparametrized}, it is possible to check that, at least formally, 
\begin{eqnarray}
\label{quantumBogliubovU}
\hat{U}_{B}^{-1} {\hat a}_{\vec{k}}^{(x,y)}\hat{U}_{B}&=&{\tilde \alpha}_k(\eta,\eta_0){\hat a}_{\vec{k}}^{(x,y)}+{\tilde \beta}_k(\eta,\eta_0){\hat b}_{\vec{k}}^{(x,y)\dagger},\nonumber \\
\hat{U}_{B}^{-1} {\hat b}_{\vec{k}}^{(x,y)\dagger}\hat{U}_{B} &=&-\overline{\tilde \beta}_k(\eta,\eta_0){\hat a}_{\vec{k}}^{(x,y)}+\overline{\tilde \alpha}_k(\eta,\eta_0){\hat b}_{\vec{k}}^{(x,y)\dagger}.
\end{eqnarray}
Hence we see that, acting with the composed operator $\hat{U}_D=\hat{U}_B\hat{U}_L$, we achieve in fact the original Bogoliubov transformation \eqref{quantumBogoliubov} of our quantum evolution.

It is obvious that $\hat{U}_L$ does not alter the vacuum $|0\rangle_D$ of our Fock representation, namely, the state with unit norm that is annihilated by all the operators ${\hat a}_{\vec{k}}^{(x,y)}$ and ${\hat b}_{\vec{k}}^{(x,y)}$. This is a consequence of the fact that $\hat{U}_L$ does not change annihilation operators into creation ones. On the other hand, using the expansion in power series of the exponential, one can also compute the action of $\hat{U}_B$ on the vacuum state $|0\rangle_D$. One gets
\begin{eqnarray}
\label{vacuumU-1}
\hat{U}_B^{-1}|0\rangle_D&=& \prod_{\vec{k}\neq\vec{\tau},(x,y)} e^{-i\big(\rho_k-c_k^{(x,y)}\big)} {\tilde \alpha}_k \bigg[ 1- \frac{{\tilde \beta}_k}{{\tilde \alpha}_k} {\hat a}_{\vec{k}}^{(x,y)\dagger} {\hat b}_{\vec{k}}^{(x,y)\dagger}\bigg] |0\rangle_D,
\\
\label{vacuumU}
\hat{U}_B|0\rangle_D&=& \prod_{\vec{k}\neq\vec{\tau},(x,y)} e^{i\big(\rho_k-c_k^{(x,y)}\big)} \overline{\tilde \alpha}_k\bigg[ 1+ \frac{\tilde{\beta}_k}{\overline{\tilde \alpha}_k} {\hat a}_{\vec{k}}^{(x,y)\dagger} {\hat b}_{\vec{k}}^{(x,y)\dagger}\bigg] |0\rangle_D.
\end{eqnarray}
Given our previous discussion, this is also the action of the complete evolution operator $\hat{U}_D$ on the vacuum. Thus, in terms of our original Bogoliubov coefficients,
\begin{equation}
\label{vacuumUD}
\hat{U}_D|0\rangle_D= \prod_{\vec{k}\neq\vec{\tau},(x,y)} e^{i\big[\rho_k-c_k^{(x,y)}-\omega_k(\eta-\eta_0)\big]} \bar{ \alpha}_k\bigg[ 1+ \frac{\beta_k}{\bar{ \alpha}_k} {\hat a}_{\vec{k}}^{(x,y)\dagger} {\hat b}_{\vec{k}}^{(x,y)\dagger}\bigg] |0\rangle_D.
\end{equation}
All these formulas are strictly rigorous if the transformed vacuum is a well-defined state on the Fock space, that is, if it has finite norm, something that happens if and only if the sequence of $\beta$-coefficents of our Bogoliubov transformation is square summable. This summability is precisely the necessary and sufficient condition for the unitary implementability of the evolution \cite{shale,dere}, implementability that we will prove in the next section. With this eventual caveat, we will now demonstrate that the transformed vacuum state \eqref{vacuumUD} is a solution to the Schr\"odinger equation of the Dirac field \eqref{schrofermion}, leaving apart the zero-modes.

Recalling the definition of the operators $\hat{\Upsilon}_F$ and $\hat{\Upsilon}_I$ in Eqs. \eqref{UpsilonF} and \eqref{UpsilonI}, and of the functions $F_k^{(\Gamma)}$ and $G_k^{(\Gamma)}$ in Eqs. \eqref{FkGamma} and \eqref{GkGamma}, we get
\begin{eqnarray}
\label{ScroHamilton}
\frac{1}{2l_0}\, \frac{\langle \hat{\Upsilon}_F+\hat{\Upsilon}_I \rangle_{\Gamma}}{\langle \hat{V}^{2/3}\rangle_{\Gamma}}\hat{U}_D|0\rangle_D&=&\sum_{\vec{k}\neq\vec{\tau},(x,y)} \Big[-F_k^{(\Gamma)} \Big( {\hat a}_{\vec{k}}^{(x,y)\dagger} {\hat a}_{\vec{k}}^{(x,y)}+  {\hat b}_{\vec{k}}^{(x,y)\dagger}  {\hat b}_{\vec{k}}^{(x,y)} \Big) \nonumber\\
&- &  i G_k^{(\Gamma)}
\Big( {\hat a}_{\vec{k}}^{(x,y)\dagger} {\hat b}_{\vec{k}}^{(x,y)\dagger} +  {\hat a}_{\vec{k}}^{(x,y)}  {\hat b}_{\vec{k}}^{(x,y)} \Big)  \Big]\hat{U}_D|0\rangle_D.
\end{eqnarray}
A simple calculation using Eq. \eqref{vacuumUD} leads then to
\begin{eqnarray}
\label{ScroHamilton2}
\frac{1}{2l_0}\,  \frac{\langle \hat{\Upsilon}_F+\hat{\Upsilon}_I \rangle_{\Gamma}}{\langle \hat{V}^{2/3}\rangle_{\Gamma}}\hat{U}_D|0\rangle_D&=&\sum_{\vec{k}\neq\vec{\tau},(x,y)} \Bigg[\Bigg(-2F_k^{(\Gamma)} \frac{\beta_k}{\bar{\alpha}_k}-i G_k^{(\Gamma)} \frac{(\beta_k)^2+(\bar{\alpha}_k)^2}{(\bar{\alpha}_k)^2} \Bigg){\hat a}_{\vec{k}}^{(x,y)\dagger} {\hat b}_{\vec{k}}^{(x,y)\dagger}\nonumber\\
&+& i G_k^{(\Gamma)}\frac{\beta_k}{\bar{\alpha}_k} \Bigg] \hat{U}_D|0\rangle_D.
\end{eqnarray}

On the other hand, taking directly the time derivative of Eq. \eqref{vacuumUD}, we obtain
\begin{eqnarray}
\label{detaGammaUD}
d_{\eta_{\Gamma}}\hat{U}_D|0\rangle_D &=& \sum_{\vec{k}\neq\vec{\tau},(x,y)} \Bigg[ i d_{\eta_{\Gamma}}\rho_k-id_{\eta_{\Gamma}}c_k^{(x,y)}-i \omega_k +\frac{d_{\eta_{\Gamma}} {\bar \alpha}_k }{ {\bar \alpha}_k}    \nonumber\\
&+ &  \frac{   {\bar \alpha}_k d_{\eta_{\Gamma}} \beta_k  - \beta_k d_{\eta_{\Gamma}} {\bar \alpha}_k  }{ ({\bar \alpha}_k) ^2  }  {\hat a}_{\vec{k}}^{(x,y)\dagger} {\hat b}_{\vec{k}}^{(x,y)\dagger}\Bigg] \hat{U}_D|0\rangle_D.
\end{eqnarray}
The quantum evolution equations \eqref{equationmotiona} and the Bogoliubov relation \eqref{quantumBogoliubov}, together with the phase redefinition performed in Eq. \eqref{tildeBogoliubov}, imply that
\begin{eqnarray}
\label{derivativa}
d_{\eta_{\Gamma}}\alpha_k&=& -iF_k^{(\Gamma)}\alpha_k-G_k^{(\Gamma)}\bar{\beta}_k,\qquad
d_{\eta_{\Gamma}}\beta_k= -iF_k^{(\Gamma)}\beta_k + G_k^{(\Gamma)}\bar{\alpha}_k, \\
d_{\eta_{\Gamma}}{\tilde \alpha}_k&=&- i \Big(F_k^{(\Gamma)}-\omega_k\Big) {\tilde \alpha}_k-G_k^{(\Gamma)}\overline{\tilde \beta}_k.
\label{derivativetildeBogoliubova}
\end{eqnarray}
In turn, we can take time derivatives in the parametrization \eqref{alphaparametrized} of ${\tilde \alpha}_k$. Substituting this parametrization in Eq. \eqref{derivativetildeBogoliubova} and identifying the results, one can prove that 
\begin{equation}
\label{derivativerho}
d_{\eta_{\Gamma}}\rho_k= \omega_k-F_k^{(\Gamma)}-G_k^{(\Gamma)} \Im{(\Delta_k)},
\end{equation}
where $\Im{(\Delta_k)}=-i(\Delta_k- {\bar\Delta}_k)/2$ is the imaginary part of $\Delta_k$. Inserting this identity and Eq. \eqref{derivativa} in our formula for the derivative of the evolved vacuum, we get
\begin{eqnarray}
\label{detaGammaUD2}
-id_{\eta_{\Gamma}}\hat{U}_D|0\rangle_D &=&
\sum_{\vec{k}\neq\vec{\tau},(x,y)} \Bigg[\Bigg(-2F_k^{(\Gamma)} \frac{\beta_k}{\bar{\alpha}_k}-i G_k^{(\Gamma)} \frac{(\beta_k)^2+(\bar{\alpha}_k)^2}{(\bar{\alpha}_k)^2} \Bigg){\hat a}_{\vec{k}}^{(x,y)\dagger} {\hat b}_{\vec{k}}^{(x,y)\dagger}\nonumber\\
&+& i G_k^{(\Gamma)}\frac{\beta_k}{\bar{\alpha}_k}- G_k^{(\Gamma)} \Im{(\Delta_k)} - d_{\eta_{\Gamma}}c_k^{(x,y)}\Bigg] \hat{U}_D|0\rangle_D.
\end{eqnarray}
So, recalling Eq. \eqref{ScroHamilton2} and employing the change of time \eqref{etaGamma}, we conclude that the evolved vacuum is indeed a solution to the Schr\"odinger equation of the Dirac field (without zero-modes). Moreover, we confirm that the fermionic Hamiltonian that generates the evolution in the time $\phi$ is $\langle \hat{\Upsilon}_F+\hat{\Upsilon}_I \rangle_{\Gamma}/(2\langle \hat{\tilde{\mathcal{H}}}_0\rangle_{\Gamma})$, modulo a backreaction contribution equal to  
\begin{equation}
\label{backreaction}
C_D^{(\Gamma)}(\phi)=-\frac{l_0\langle \hat{V}^{2/3}\rangle_{\Gamma}}{\langle \hat{\tilde{\mathcal H}}_0 \rangle_{\Gamma}} \sum_{\vec{k}\neq\vec{\tau},(x,y)} \Big[G_k^{(\Gamma)} \Im{(\Delta_k)} + d_{\eta_{\Gamma}}c_k^{(x,y)}\Big].
\end{equation}

\section{Unitarity and backreaction considerations}

The unitarity of the evolution operator introduced in the previous section can be dilucidated by checking whether the $\beta$-coefficients of the corresponding Bogoliubov transformation, that provides the change in time of the creation and annihilation operators, are square summable or not at all instants in the time interval under consideration. Assuming the finiteness of those coefficients, the summability depends only on the asymptotic behavior for large values of $\omega_k$, where the contribution of an infinite number of fermionic modes can result in a divergence. 

Let us start by considering the functions $F_k^{(\Gamma)}$ and $G_k^{(\Gamma)}$, that encode the information about the expectation values on the state $\Gamma$ of the homogeneous geometry that is relevant for the dynamics of the fermion field. We use their expressions \eqref{FkGamma} and\eqref{GkGamma} and the definition of $\breve{\xi}_k$ in Eq. \eqref{xik}, rewritten in terms of the volume $V=4\pi\sigma e^{3\breve{\alpha}}/3$, namely
\begin{equation}
\label{xikV}
\breve{\xi}_k(\hat{V})=\sqrt{\omega_k^2+\frac{M^2 \hat{V}^{2/3}}{l_0^{2}}}.
\end{equation}
If we then introduce the spectral decomposition of $\Gamma$ associated with the operator $\hat{V}$, and call $\breve{V}$ the studied eigenvalue, we can express $\breve{\xi}_k(\hat{V})$ in a series expansion in powers of $\hat{V}^{2/3}$ at least for $\omega_k > M \breve{V}^{1/3}/l_0$, something that happens in the ultraviolet region of infinitely large $\omega_k$. In this way, one gets the following expansions of the functions $F_k^{(\Gamma)}$ and $G_k^{(\Gamma)}$, which provide in fact their asymptotic Laurent series in powers of $\omega_k$:
\begin{equation}
\label{FkGammaseries}
F_k^{(\Gamma)}=\omega_k- \sum_{n=1}^{\infty}  (-1)^{n} \frac{M^{2n}}{l_0^{2n}\omega_k^{2n-1}} \frac{(2n-3)!!}{2^n n!} W_n^{(\Gamma)},\qquad
W_n^{(\Gamma)}= \frac{\langle \hat{V}^{2(n+1)/3}\rangle_{\Gamma}} {\langle \hat{V}^{2/3}\rangle_{\Gamma}} ,
\end{equation}
and
\begin{eqnarray}
\label{GkGammaseries}
G_k^{(\Gamma)}&=& \sum_{n=0}^{\infty} (-1)^n \frac{M^{2n+1}}{l_0^{2(n+1)}\omega_k^{2n+1}} \frac{\lambda_n^{(\Gamma)}}{2^{n+1}}  ,\\
\label{OmeganGamma}
\lambda_n^{(\Gamma)}&=& \sum_{m=0}^{n} \frac{(2m-1)!!(2n-2m-1)!!}{m! (n-m)!} \frac{ \langle  \hat{V}^{(4m+1)/6}  \hat{\Lambda}_0 \hat{V}^{[4(n-m)+1]/6} \rangle_{\Gamma} }{\gamma \langle \hat{V}^{2/3} \rangle_{\Gamma} },
\end{eqnarray}
where $n!!$ is the double factorial of the integer $n$, identified with the unity if $n\leq 0$.

Substituting these formulas in Eq. \eqref{HkGamma}, and expanding the square roots and denominators, one can obtain an asymptotic series for $H_k^{(\Gamma)}$. We give here only the leading orders:
\begin{equation}
\label{HkGammaseries}
H_k^{(\Gamma)}= \frac{M}{ l_0} \sqrt{ W_1^{(\Gamma)}  }  \left[ -i + \frac{1}{4\omega_k} \left(\ln{  W_1^{(\Gamma)}   } \right)^{\prime\,}\, \right]
- \frac{M}{2 l_0^2\omega_k} \lambda_0^{(\Gamma)}+{\mathcal{O}}(\omega_k^{-2}),
\end{equation}
where, with our notation,
\begin{equation}
\label{Omega0Gamma}
 W_1^{(\Gamma)} =\frac{\langle \hat{V}^{4/3}\rangle_{\Gamma}} {\langle \hat{V}^{2/3}\rangle_{\Gamma}},\qquad \lambda_0^{(\Gamma)}=\frac{ \langle  \hat{V}^{1/6}  \hat{\Lambda}_0 \hat{V}^{1/6} \rangle_{\Gamma} }{\gamma \langle \hat{V}^{2/3} \rangle_{\Gamma} }.
\end{equation}
This formula allows us to get the asymptotic behavior of $\zeta_k$. In doing so, apart from employing the asymptotic series of $H_k^{(\Gamma),0}$, we have to express the denominator in Eq. \eqref{zeta} as a power series in the inverse of $\omega_k$. The required series expansion is possible at least for $\omega_k>|(\ln{H_k^{(\Gamma)}})^{\prime}_0|/2$, or, approximating the right-hand side of this inequality by means of Eq. \eqref{HkGammaseries}, for $\omega_k>|(\ln{ W_1^{(\Gamma)} })^{\prime}_0|/4$, condition that holds in the studied asymptotic region. We get at leading orders
\begin{equation}
\label{zetaseries}
\zeta_k=\frac{M}{2l_0\omega_k} \sqrt{ W_1^{(\Gamma),0} }  
-i\frac{M}{4l_0^2\omega_k^2} \lambda_0^{(\Gamma),0}+{\mathcal{O}}(\omega_k^{-3}),
\end{equation}
where again the superscript $0$ means evaluation at the initial time. In addition, we can derive the asymptotic expansion of the integral of $\Lambda_k^l$ using Eqs. \eqref{uele}, \eqref{Lambdaasymptotics}, and \eqref{HkGammaseries}. For our discussion, we will only need 
\begin{eqnarray}
\label{Lambdaseries}
\int_{\eta_0}^{\eta} \left[\Lambda_k^1(\eta_{\Gamma})-\bar{\Lambda}_k^2(\eta_{\Gamma})\right]   d\eta_{\Gamma} &=& - \frac{1}{2\omega_k} \left[ \Big( \ln{H_k^{(\Gamma)} } \Big)^{\prime }-\Big(\ln{H_k^{(\Gamma)} }\Big)^{\prime}_0 \right]+ {\mathcal{O}}(\omega_k^{-2})\nonumber\\
&  =& -\frac{1}{4\omega_k} \left[ \left(\ln{  W_1^{(\Gamma)} } \right)^{\prime}- \left(\ln{  W_1^{(\Gamma)} } \right)^{\prime}_0 \,  \right]+ {\mathcal{O}}(\omega_k^{-2}).
\end{eqnarray}
Note that, without entering the asymptotic region of extremely large $\omega_k$, each of the two contributions to the dominant term is smaller than the unity in absolute value, and then we expect that the whole quantity will be small, if we have that, both at the considered value $\eta$ of the conformal time and at the initial value $\eta_0$,
\begin{equation}
\label{condiasymptotic2}
\omega_k>\frac{1}{4}\left|\left(\ln{ W_1^{(\Gamma)} } \right)^{\prime}\right|.
\end{equation} 
In particular, at the initial time, this condition guarantees the requirement $\omega_k>|(\ln{W_1^{(\Gamma)}})^{\prime}_0|/4$ that we had found in our discussion above.

Let us discuss now the asymptotic behavior of $f_{1,k}^{(\Gamma)}$ and $f_{2,k}^{(\Gamma)}$. We can calculate their asymptotic series using the expansion \eqref{FkGammaseries} and expressions \eqref{quantumDEHfs}. We obtain
\begin{eqnarray}
\label{f1series}
f_{1,k}^{(\Gamma)}&=& \frac{M}{2l_0\omega_k} \sqrt{ W_1^{(\Gamma)} }+ {\mathcal{O}}(\omega_k^{-3}),
\\
\label{f2series}
f_{2,k}^{(\Gamma)}&=& 1- \frac{M^2}{8l_0^2\omega_k^2} W_1^{(\Gamma) }+ {\mathcal{O}}(\omega_k^{-4}).
\end{eqnarray}

At last, we can compute the asymptotic expansion of the $\beta$-coefficient of the dynamical Bogoliubov transformation by inserting  in the formula \eqref{beta} all the pieces about the asymptotic behavior that we have accumulated in this section. A careful calculation leads to
\begin{equation}
\label{betaseries}
\beta_k= i\frac{M}{4l_0^2\omega_k^2} \left[ \lambda_0^{(\Gamma),0} e^{-i\omega_k (\eta-\eta_0)}- \lambda_0^{(\Gamma)}e^{i\omega_k (\eta-\eta_0)}\right]+{\mathcal{O}}(\omega_k^{-3}).
\end{equation}
In this way, we reach the important conclusion that $\beta_k$ is of the asymptotic order of $\omega_k^{-2}$. Since the degeneracy (i.e., the number of tuples $\vec{k}$ with the same value of $\omega_k$) is at most of the order of $\omega_k^2$ in the ultraviolet limit under consideration, it follows that the sequence formed by the $\beta$-coefficients is indeed square summable. Even if we expected this result, based on our choice of Fock representation and on the strategy followed in the hybrid approach, the inclusion of quantum fluctuations in the background casted shadows over the unitary implementability of the evolution. We see that, definitively, the quantum dynamics of the nonzero fermionic modes is unitary. This closes the only point left open in the proof that the evolved vacuum $\hat{U}_D|0\rangle_D$ is a solution to the Sch\"odinger equation of the nonzero-modes of the Dirac field.

In addition, it is well-known that the number of particles produced out of the vacuum in the evolution, as perceived by the original vacuum (i.e., according to its notion of particles) coincides in fact with the sum of the square norm of the $\beta$-coefficients \cite{DEH,wald}. This quantity is also the number of antiparticles created in the evolution, since particles and antiparticles appear in pairs. Therefore, the production of particles (or, strictly speaking, of pairs of particles and antiparticles) is finite with our choice of creation and annihilation operators for the Dirac field, as it was already shown in the geometrodynamical case in Ref. \cite{DEH}. Actually, we see from Eq. \eqref{betaseries} that the contribution of modes with large $\omega_k$ is proportional to the square mass of the fermion field, and hence really small, taking into account that this mass is typically insignificant, e.g. around $10^{-23}$ for the electron in Planck units. This property, together with the decaying behavior of the production of particles as $\omega_k^{-4}$ for large $\omega_k$, guarantees that the fermionic part of the state does not depart much from the vacuum for modes that are well inside the cosmological horizon when semiclassical trajectories are chosen for the expectation values of the homogeneous geometry. Therefore, our results prove to be compatible with the expected behavior of fermion fields in the low-curvature regions of the spacetime. Besides, the square norm of $\beta_k$ at dominant asymptotic order is proportional to the sum of the squares of $\lambda_0^{(\Gamma)}$ and $\lambda_0^{(\Gamma),0}$, both of which are real, plus an oscillating term. The contribution of this last term to the particle production will be negligible compared with the other two, since the sum over modes will average the huge asymptotic oscillations. We also notice that the definition of $\lambda_0^{(\Gamma)}$ in Eq. \eqref{Omega0Gamma} involves the operator $\hat{\Lambda}_0$. This operator is a modified version of $\hat{\Omega}_0$ in which the length of the holonomies has been doubled. For states that are highly peaked on genuine classical trajectories, in regions where general relativity holds, one expects $\lambda_0^{(\Gamma)}$ to be approximately equal to the derivative of the scale factor with respect to the conformal time, apart from a multiplicative constant. But in LQC, beyond those regions, it provides the quantum value of the Hubble constant, which is known to vanish in the big bounce on effective trajectories. These trajectories depart from general relativity, as we have commented, when the energy density approaches the Planck density \cite{APS1}, even though there still exist quantum states peaked on them. Therefore, if we choose the initial time at the big bounce and we consider a state that, at least around $\eta_0$, is peaked on an effective trajectory, the contribution of the initial value of $\lambda_0^{(\Gamma)}$, namely $\lambda_0^{(\Gamma),0}$, should vanish (or be negligible). In this sense, LQC is able to provide initial conditions that minimize the production of fermionic particles. Note that a similar situation would not be possible, for instance, in geometrodynamics, because the Hubble constant will never vanish in that approach. 

As for the production of particles in modes that do not belong to the ultraviolet region, one can follow an analysis similar to that presented in Ref. \cite{DEH}. Actually, the expansions and approximations that we have carried out in the deduction of Eq. \eqref{betaseries} are valid if condition \eqref{condiasymptotic2} is verified at the conformal times $\eta$ and $\eta_0$, and if $\omega_k >M\langle \hat{V}^{1/3}\rangle_{\Gamma}/l_0$, where we have approximated the action of the operator $\hat{V}^{1/3}$ by its expectation value. On quantum states that are peaked on effective trajectories, the right-hand side of Eq. \eqref{condiasymptotic2} can be estimated, apart from an irrelevant multiplicative number, as the Hubble parameter multiplied by the scale factor. Hence, in the case of LQC and choosing the bounce as the initial instant, the condition at $\eta_0$ reduces to the trivial demand that $\omega_k>0$, which can be ignored. Again, this would not happen with other approaches to the quantization of the homogeneous geometry. In total, we expect that our restrictions imply that $\omega_k>\omega_k^0$, where $\omega_k^0$ is the larger of $M\langle \hat{V}^{1/3}\rangle_{\Gamma}/l_0$ and $|(\ln{W_1^{(\Gamma)}})^{\prime}|/4$. For modes that do not satisfy this inequality, the particle production should be of the order of unity per mode, adding to a total quantity proportional to the cube of $\omega_k^0$, where we have taken into account the degeneracy of the modes. Let us emphasize that the physically important result is the finiteness of the number of particles, in spite of the presence of an infinite number of modes. This means that the ultraviolet modes do not depart considerably from their vacuum, and thanks to this fact they make very little contribution to the particle production.

Let us complete our asymptotic analysis by considering the behavior or some  additional quantities, related to the backreaction contribution of the fermionic field. A calculation similar to that explained for $\beta_k$, but now using Eq. \eqref{alpha}, confirms that
\begin{equation}
\label{alphaseries}
\alpha_k = e^{-i\omega_k (\eta-\eta_0)} + {\mathcal{O}}(\omega_k^{-1}),
\end{equation}
as we anticipated at the beginning of the previous section. If we make use of this asymptotic expression and of Eq. \eqref{betaseries}, and recall the parametrization \eqref{alphaparametrized} of the phase-shifted Bogoliubov coefficients, we can check that $\rho_k={\mathcal{O}}(\omega_k^{-1})$, whereas $\tilde{\beta}_k=e^{-i\omega_k (\eta-\eta_0)} \beta_k$ coincides with  $-\Delta_k$ up to subdominant terms of order $\omega_k^{-3}$ or less, so that
\begin{equation}
\label{Deltaseries}
\Im(\Delta_k)= \frac{M}{4l_0^2\omega_k^2} \left\{  \lambda_0^{(\Gamma)}-\lambda_0^{(\Gamma),0} \cos{\left[2\omega_k (\eta-\eta_0)\right]}\right\}+{\mathcal{O}}(\omega_k^{-3}).
\end{equation}
Multiplying this identity by $G_k^{(\Gamma)}$ and using the expansion \eqref{GkGammaseries}, we get 
\begin{equation}
\label{backseries}
G_k^{(\Gamma)}\Im(\Delta_k)= \frac{M^2}{8l_0^4\omega_k^3} \lambda_0^{(\Gamma)} \left\{\lambda_0^{(\Gamma)}-\lambda_0^{(\Gamma),0} \cos{\left[2\omega_k (\eta-\eta_0)\right]}\right\}+{\mathcal{O}}(\omega_k^{-4}).
\end{equation}
The sum over all modes of the subdominant terms of order $\omega_k^{-4}$ in this expression converges, because the degeneracy grows asymptotically at most as a function of order $\omega_k^2$. Hence, in the backreaction contribution \eqref{backreaction} of the nonzero-modes of the Dirac field, the only possible divergences arising from $G_k^{(\Gamma)}\Im(\Delta_k)$ may come from its dominant term, of order $\omega_k^{-3}$. Actually, the oscillating part proportional to $\lambda_0^{(\Gamma),0}$ can be ignored in LQC if we choose the initial time at the bounce and the state $\Gamma$ to be sufficiently peaked on an effective trajectory around $\eta_0$, as we have argued in the discussion of the particle production. In this case, the only possible divergent contribution would be that of $M^2 (\lambda_0^{(\Gamma)})^2 /(8l_0^4\omega_k^3)$. The presence of this divergence requires a regularization process, that can be incorporated in our discussion by means of a suitable choice of the phase $c_k^{(x,y)}$ in the fermionic Hamiltonian. The divergence is absorbed with a choice of the form
\begin{equation}
\label{backreactionphase}
c_k^{(x,y)}= -\frac{M^2}{8l_0^4\omega_k^3} \int_{\eta_0}^{\eta}  \Big(\lambda_0^{(\Gamma)}\Big)^2 +{\mathcal{O}}(\omega_k^{-4}).
\end{equation}
We notice that the dominant term is independent of the considered pair $(x,y)$ and vanishes at the initial time, since the same happens with the divergent part of \eqref{backseries}.

This contrasts with the situation found in Ref. \cite{DEH}, where the divergent contribution of each fermionic mode was shown to be proportional to $\omega_k$, much worse than the behavior ${\mathcal{O}}(\omega_k^{-3})$ found here. The improvement in the quantum theory with respect to these divergences in absence of regularization must be attributed mainly to our selection of Fock representation for the nonzero-modes of the Dirac field, instead of using the same holomorphic representation adopted in the mentioned work. On the other hand, arguments of the kind explained in that reference lead to the expectation that the fermionic backreaction, after regularization, should have negligible effects for the typical small values of the fermion mass (even the divergent term is proportional to the square mass). Moreover, in the next section we will present further comments pointing out to the possibility that this backreaction can be made finite without regularization, just by adapting in an optimal way our selection of Fock quantization, while remaining in the same unitary equivalence class of the choice discussed here.

\section{Discussion}

In this work, we have discussed the quantization of a Dirac field coupled to a perturbed flat FLRW spacetime with a massive scalar field in the framework of LQC. For mathematical convenience, we have assumed compact spatial sections. This hypothesis should not have physically relevant consequences in cosmology, at least if the compactification scale is much larger than that corresponding to the cosmological horizon. In our quantization, the geometry and the scalar matter content have been treated quantum mechanically as well. Moreover, we have allowed the presence of scalar and tensor perturbations in the quantum system. Vector perturbations do not play any physical role, since they are gauge degrees of freedom in our model and can be ignored. In practice, the Dirac field has also been treated as a perturbation, inasmuch as any possible contribution of the fermionic zero-modes has been supposed small, if the spin structure permits that such modes exist. We have truncated the Dirac-Einstein action, with the coupling to the scalar field, at quadratic order in the perturbations. For the modes of the Dirac field, the truncation has no effect, since their contribution to the action is already quadratic. Our analysis can be considered an extension to LQC of the work of D'Eath and Halliwell \cite{DEH} in quantum geometrodynamics, with certain additional distinctive features that will be pointed out in the following discussion.

We have adopted a hybrid quantization strategy, with different kind of representations for the sector of homogeneous degrees of freedom of our system and for its inhomogeneities. The quantum representation of the total system is a direct product, but the system is highly nontrivial because the distinct sectors of variables are coupled by constraints, arising from those of general relativity and from the gauge symmetries. For the homogeneous sector, the scalar perturbations, and the tensor ones, it is possible to introduce a canonical transformation that disentangles the constraints that have a genuine perturbative nature. This leads to a set of variables for the perturbations that includes those constraints, some suitable conjugate momenta, and canonical pairs of gauge invariants (namely, variables that commute with the gauge transformations generated by the perturbative constraints). Besides, this transformation provides homogeneous variables that retain the canonical structure, not only among them, but also with respect to the perturbations \cite{hybr-ref}. After the transformation, it is almost straightforward to deal with the perturbative constraints \`a la Dirac. Physical states depend only on the homogeneous variables, the gauge-invariant scalar and tensor perturbations, and the fermionic modes. The relevant constraint left in the system is the zero-mode of the Hamiltonian constraint, in which the homogeneous constraint that persists in absence of perturbations appears modified by quadratic perturbative contributions. 

According to our hybrid approach, we have then passed to a convenient Fock description of the Dirac field. For its nonzero-modes, we have employed the same type of creation and  annihilation variables adopted in Ref. \cite{DEH}. This choice belongs to a privileged family of unitarily equivalent Fock representations for the fermionic degrees of freedom. When the background spacetime is regarded as a classical entity, this family is picked out by the criterion of a unitarily implementable dynamics, together with the invariance of the vacuum under the spatial symmetries of the system and the spin rotations generated by the helicity \cite{uf-flat}, adopting a convention of particles and antiparticles that connects smoothly with the standard one in the massless case. In contrast to the analysis of Ref. \cite{DEH} where, at the end of the day, a holomorphic representation was selected for the Dirac field, we have truly performed the change to these creation and annihilation variables in our Hamiltonian treatment. Since this change depends on the configuration variable that describe the homogeneous geometry (let us say its scale factor), we have had to complete it into a canonical transformation for the entire system, at our order of perturbative truncation. This has two effects. First, it requires a modification of the canonical momentum of the homogeneous geometry, incorporating quadratic contributions of the fermions. This modification allows us to retain a symplectic (canonical) structure in our system, and to progress in the discussion with a neat correspondence between our canonical variables and the metric and matter fields. Second, it alters the global Hamiltonian constraint, since the transformation involves time-varying variables. As a consequence, the final global Hamiltonian constraint for our hybrid quantization differs from that of Ref. \cite{DEH}. Our Hamiltonian dictates the evolution of the system after a splitting of the degrees of freedom in which a specific part of the dynamics of the Dirac field is attributed to the homogeneous geometry and another to the selected creation and annihilation variables, rather than to the fermionic variables associated with the holomorphic representation used by D'Eath and Halliwell. 

In the hybrid quantization of this system, we have introduced a Born-Oppenheimer ansatz, searching for quantum physical states that present a separated dependence on the homogeneous geometry, the gauge-invariant scalar perturbations, the tensor perturbations, and the fermionic degrees of freedom. In this ansatz, the role of the homogeneous scalar field is that of an internal time. The corresponding wave function of the homogeneous geometry has been constructed as a state evolving unitarily in terms of that internal time. The generator of this unitary evolution is assumed to be perturbatively close to that of the unperturbed model. We have then identified the conditions necessary to ignore transitions of the homogeneous geometry mediated by the global, zero-mode of the Hamiltonian constraint. In this way we have arrived at a master constraint equation that is quadratic in the momentum of the scalar field and in all the perturbative elementary operators (of scalar, tensor, or fermionic nature), and where the homogeneous geometry is incorporated only via expectation values. From this master constraint, and some very mild assumptions about the contributions of the perturbations to the momentum of the homogeneous scalar field, one can derive, for instance, the quantum counterpart of the Mukhanov-Sasaki equations for the gauge-invariant scalar perturbations \cite{hybr-ref}, or the quantum dynamics of the fermionic variables. Alternatively, we have specified conditions so that, from this master constraint equation, one can extract Schr\"odinger equations for the different perturbations of the system. These Schr\"odinger equations use again the homogeneous scalar field as a natural time for the quantum evolution. The procedure to arrive to these equations also differs in some fundamental aspects from the discussion presented in Ref. \cite{DEH}. There the authors recurred to a semiclassical approximation, starting from an action that was a Hamilton-Jacobi solution of the homogeneous model and defining a notion of time evolution in terms of the projection in the direction of the gradient of this action, all this performed in absence of a Hilbert space and an inner product for the homogeneous geometry. In our case, the Hilbert space and the inner product are those of LQC, and the expectation values that appear in the master constraint equation are rigorously defined and capture the quantum behavior of the wave function of the homogeneous geometry, without the need of assuming semiclassical trajectories. Nevertheless, let us clarify that many aspects of our treatment can be generalized to other approaches to the quantization of the homogeneous geometry, different from LQC, along the lines that we have sketched at the end of Sec. V. On the other hand, and partly related to the issue of time that we were commenting, we have also introduced a conformal time \eqref{etaGamma} that depends on the particular state considered for the homogeneous geometry. This time is well-defined in our quantization, at the level of our perturbative truncation. In geometrodynamics, however, any possible semiclassical counterpart of the definition of this time would be problematic around the big bang, since the numerator of Eq. \eqref{etaGamma} would vanish there, given that the cosmological singularity is not eluded in the semiclassical trajectories.

For the fermionic nonzero-modes, we have analyzed in detail the quantum dynamics. The evolution equations retain the effects of the quantization of the homogeneous geometry by means of the presence of expectation values, that replace the role played by functions of the background in ordinary quantum field theory in curved spacetimes. At this point, several comments are in order. 

The dynamical equations for all the infinite tower of modes of the (gauge-invariant) scalar and tensor perturbations depend only on a finite number of expectation values. In full contrast, in the case of the Dirac field, its dynamics depends on different expectation values for each of the modes, leading to an infinite sequence of them. This has radical implications for the philosophy that has been put forward in the dressed metric approach, where the standard interpretation is that a limited number of expectation values must suffice to characterize the dressed metric, encoding essentially the same information that determines the homogeneous solutions to the effective description of LQC \cite{dressed1,dressed2,dressed3}. We also emphasize that the  expectation values that enter the dynamical equations of the sequence of creation and annihilation variables of the Dirac field explore the quantum dependence of the wave function of the homogeneous geometry in an infinite number of algebraic powers of the volume, but they always depend in the same operational way on the conjugate momentum. 

In spite of the inclusion of these quantum effects on the geometry, and the subsequent replacement of functions of the background by expectation values in the dynamical equations of the fermionic variables, the resulting dynamics determines a Bogoliubov transformation that is indeed implementable as a unitary transformation in our hybrid quantization. This result is reassuring and permits a rigorous connection between LQC and quantum filed theory in curved backgrounds. We have found the unitary operator that implements the dynamical evolution and proved that it is generated by a fermionic Hamiltonian that, as expected, coincides with the Hamiltonian that appears in the Schr\"odinger equation deduced for the system. In particular, we have constructed a solution to this equation, which describes the evolution of the (nonzero-modes part of the) fermionic vacuum. The construction is exact: The transformed vacuum satisfies the Schr\"odinger equation without further approximations.

The identification of a notion of quantum dynamics for the fermionic degrees of freedom that is unitarily implementable, and its realization in terms of a specific operator and its corresponding Hamiltonian, guarantees the quantum coherence of the evolution in what refers to the associated concept of particles and antiparticles. In other words, we have been able to split the evolution of the Dirac field into a part that varies with the homogeneous geometry and another for which the dynamics can be implemented as a unitary transformation in the quantum theory (at least in the range of validity of the Schr\"odinger equation). Even if the geometry is a quantum dynamical entity as well, and in general is far from being stationary, the unitarity of the fermionic dynamics preserves the coherence, and hence the quantum information, about the particles and antiparticles described by the creation and annihilation variables that we have picked out in our Fock quantization.

In particular, the unitarity immediately ensures that the particle production, or equivalently the creation of pairs of particles and antiparticles, is finite. For modes with $\omega_k$ smaller or of the order of the fermion mass and the Hubble parameter, we have recurred to arguments similar to those explained in Ref. \cite{DEH} to estimate that the number of particles per unite volume will be, roughly speaking, proportional to the cube of the larger of the two considered quantities, that in inflationary scenarios is typically the Hubble scale. Much more importantly, for any mode with larger $\omega_k$, and especially those deep inside the ultraviolet sector, the production is insignificant and, in general, proportional to the square fermion mass. The number is small enough as to have a convergent sum when all modes are considered. In this sense, we can consider that the modes in the ultraviolet sector, which do not cross the cosmological horizon, do not depart significantly from their vacuum state in the quantum evolution. Let us also comment that, for modes which do not really belong to the region of asymptotically large $\omega_k$, one could modify the vacuum state following prescriptions like, for instance, that introduced recently by Mart\'{\i}n de Blas and Olmedo \cite{hybr-pred}. That prescription is characterized precisely by minimizing the particle production in the evolution while not affecting in a relevant way the physical behavior of the vacuum in the ultraviolet sector. 

We have also investigated the issue of the backreaction contribution of the fermions to the master constraint equation, which includes not only the homogeneous sector, but also the scalar and tensor perturbations. This backreaction contribution is identified as the fermion-independent part of the quantum Hamiltonian for the Dirac field, once we have adopted the normal ordering corresponding to our choice of Fock representation. We have shown that this backreaction needs regularization, as it was already discussed in Ref. \cite{DEH}, but the situation is now much better than the one found in the geometrodynamical analysis. The divergent individual contributions of each mode in the ultraviolet sector are now of the order ${\mathcal{O}}(\omega_k^{-3})$, instead of order ${\mathcal{O}}(\omega_k)$. Actually, any contribution of order ${\mathcal{O}}(\omega_k^{-3+\varepsilon})$, with arbitrary $\varepsilon>0$, is summable, and hence leads to a finite backreaction effect. We have postponed to this point of the discussion an important comment. As we have explained, we still have certain freedom in the choice of Fock description for our Dirac field while respecting the criterion of unitary dynamics and symmetry invariance. This freedom corresponds to a privileged family of unitarily equivalent quantizations, which do not only allow for equivalent complex structures, but also for slightly different dynamics \cite{uf-flat}. All of these dynamics are unitarily related, and the change from one to another amounts to a distinct splitting of the dependence of the Dirac field into components that depend on the homogeneous geometry and fermionic variables. We have also explained that these different dynamics have a different global Hamiltonian constraint associated with them. It should then be obvious that we still have freedom left to improve the behavior of the backreaction contribution by optimizing our choice of creation and annihilation variables for the nonzero-modes of the Dirac field, a choice that, being dependent on the homogeneous geometry, captures the available freedom in the selection of dynamics and of complex structure for the corresponding canonical anticommutation relations. The use of this freedom to deal with issues like the backreaction without recurring to regularization in LQC will be explored in future works. In this sense, our study has to be regarded as a first step towards the rigorous consideration of the backreaction in the quantization of cosmological perturbations. The system investigated here has the advantage of presenting a well understood transition from the quantum cosmology regime to the quantum field theory regime in a quantum corrected, curved spacetime, as we have shown, something that comes combined with the possibility of performing analytical computations and estimations with a great deal of accuracy and control in the quantum theory. 

\acknowledgments

The authors thank J. Cortez, D. Mart\'{\i}n de Blas, J. Olmedo, and J.M. Velhinho for enlightening discussions. This work was supported by the Project. No. MINECO FIS2014-54800-C2-2-P from Spain, and by the Portuguese Grant No. UID/FIS/04434/2013 from the Funda\c ca\~o para a Ci\^encia e Tecnologia (FCT). M. M-B.  is supported by a FCT Research contract, with Reference No. IF/00431/2015.

\end{document}